\documentclass[12pt]{emulateapj}

\usepackage{natbib}
\usepackage{graphicx}
\usepackage{psfig} 
\usepackage{color}

\newcommand{\kms}{\ifmmode {\rm km~s}^{-1} \else km~s$^{-1}$\fi}
\newcommand{\ergs}{\ifmmode {\rm erg~s}^{-1} \else erg~s$^{-1}$\fi}
\newcommand{\ergscm}{\ifmmode {\rm erg~s}^{-1} \else erg~s$^{-1}$~cm$^{-2}$\fi}
\newcommand{\Msun}{\ifmmode {\rm M}_{\odot} \else $M_{\odot}$\fi }
\newcommand{\Lsun}{\ifmmode {\rm L}_{\odot} \else L$_{\odot}$\fi}
\newcommand{\qo}{\ifmmode q_{\rm o} \else $q_{\rm o}$\fi}
\newcommand{\Ho}{\ifmmode H_{\rm o} \else $H_{\rm o}$\fi}
\newcommand{\ho}{\ifmmode h_{\rm o} \else $h_{\rm o}$\fi}

\newcommand{\gtsim}{\raisebox{-.5ex}{$\;\stackrel{>}{\sim}\;$}}
\newcommand{\vFWHM}{\ifmmode v_{\mbox{\tiny FWHM}} \else
                    $v_{\mbox{\tiny FWHM}}$\fi}
\newcommand{\CCF}{\ifmmode F_{\it CCF} \else $F_{\it CCF}$\fi}
\newcommand{\ACF}{\ifmmode F_{\it ACF} \else $F_{\it ACF}$\fi}
\newcommand{\Halpha}{\ifmmode {\rm H}\alpha \else H$\alpha$\fi}
\newcommand{\Hbeta}{\ifmmode {\rm H}\beta \else H$\beta$\fi}
\newcommand{\Hgamma}{\ifmmode {\rm H}\gamma \else H$\gamma$\fi}
\newcommand{\Hdelta}{\ifmmode {\rm H}\delta \else H$\delta$\fi}
\newcommand{\Lya}{\ifmmode {\rm Ly}\alpha \else Ly$\alpha$\fi}
\newcommand{\Lyb}{\ifmmode {\rm Ly}\beta \else Ly$\beta$\fi}
\newcommand{\HeI}{\ifmmode {\rm He}\,{\sc i}\,\lambda5876 \else 
	          He\,{\sc i}\,$\lambda5876$\fi}
\newcommand{\HeII}{\ifmmode {\rm He}\,{\sc ii}\,\lambda4686 \else 
	           He\,{\sc ii}\,$\lambda4686$\fi}
\newcommand{\heii}{He\,{\sc ii}}

\newcommand{\ciii}{\ifmmode {\rm C}\,{\sc iii}] \else C\,{\sc iii}]\fi}
\newcommand{\civ}{C\,{\sc iv}}
\newcommand{\siiv}{Si\,{\sc iv}}
\newcommand{\aliii}{Al\,{\sc iii}}
\newcommand{\nv}{N\,{\sc v}}
\newcommand{\mgii}{Mg\,{\sc ii}}

\newcommand{\mbh}{$M_{\rm BH}$\ }

\newcommand{\absmag}{$-27.69$} 
\newcommand{\density}{$n_e$~\gtsim~3.9~$\times~10^5$~cm$^{-3}$}

\shorttitle{Rapid \civ \ BAL Variability}
\shortauthors{Grier et al.}

\begin{document}

\title{The Sloan Digital Sky Survey Reverberation Mapping Project: Rapid \boldmath{\civ} Broad Absorption Line Variability}

\author{C.~J.~Grier\altaffilmark{1},
P.~B.~Hall\altaffilmark{2},
W.~N. Brandt\altaffilmark{1,3}, 
J.~R.~Trump\altaffilmark{1,4},
Yue~Shen\altaffilmark{4,5,6}, 
M.~Vivek\altaffilmark{7},
N.~Filiz~Ak\altaffilmark{8}, 
Yuguang~Chen\altaffilmark{9},
K.~S.~Dawson\altaffilmark{7},
K.~D.~Denney\altaffilmark{10,11}, 
Paul~J.~Green\altaffilmark{12}
Linhua~Jiang\altaffilmark{6}, 
C.~S.~Kochanek\altaffilmark{10,11}, 
Ian~D.~McGreer\altaffilmark{13},
I.~P{\^a}ris\altaffilmark{14},
B.~M.~Peterson\altaffilmark{10,11}, 
D.~P.~Schneider\altaffilmark{1},
Charling~Tao\altaffilmark{15,16}, 
W.~M.~Wood-Vasey\altaffilmark{17}, 
Dmitry~Bizyaev\altaffilmark{18},
Jian~Ge\altaffilmark{19},
Karen~Kinemuchi\altaffilmark{18},
Daniel~Oravetz\altaffilmark{18},
Kaike~Pan\altaffilmark{18},
Audrey~Simmons\altaffilmark{18}
}

\altaffiltext{1}{Dept.\ of Astronomy and IGC, The Pennsylvania State University, 525 Davey Laboratory, University Park, PA 16802}
\altaffiltext{2}{Department of Physics and Astronomy, York University, Toronto, ON M3J 1P3, Canada}
\altaffiltext{3}{Department of Physics, The Pennsylvania State University, University Park, PA 16802, USA}
\altaffiltext{4}{Hubble Fellow} 
\altaffiltext{5}{Carnegie Observatories, 813 Santa Barbara Street, Pasadena, CA 91101, USA} 
\altaffiltext{6}{Kavli Institute for Astronomy and Astrophysics, Peking University, Beijing 100871, China}
\altaffiltext{7}{Department of Physics and Astronomy, University of Utah, Salt Lake City, UT 84112, USA} 
\altaffiltext{8}{Faculty of Sciences, Department of Astronomy and Space Sciences, Erciyes University, 38039 Kayseri, Turkey}
\altaffiltext{9}{Department of Astronomy, School of Physics, Peking University, Beijing 100871, China}
\altaffiltext{10}{Department of Astronomy, The Ohio State University, 140 W 18th Ave, Columbus, OH 43210, USA}
\altaffiltext{11}{Center for Cosmology and AstroParticle Physics, The Ohio State University, Columbus, OH 43210, USA}
\altaffiltext{12}{Harvard-Smithsonian Center for Astrophysics, 60 Garden Street, Cambridge, MA 02138, USA}
\altaffiltext{13}{Steward Observatory, The University of Arizona, 933 North Cherry Avenue, Tucson, AZ 85721-0065, USA}
\altaffiltext{14}{INAF-Osservatorio Astronomico di Trieste, Via G. B. Tiepolo 11, I-34131 Trieste, Italy}
\altaffiltext{15}{Centre de Physique des Particules de Marseille, Aix-Marseille Universite, CNRS /IN2P3, 163, avenue de Luminy,
Case 902, F-13288 Marseille Cedex 09, France}
\altaffiltext{16}{Tsinghua Center for Astrophysics, Tsinghua University, Beijing 100084, China}
\altaffiltext{17}{PITT PACC, Department of Physics \&Astronomy, University of Pittsburgh, 3941 O'Hara Street, Pittsburgh, PA 15260}
\altaffiltext{18}{Apache Point Observatory and New Mexico State University, P.O. Box 59, Sunspot, NM, 88349-0059, USA}
\altaffiltext{19}{Department of Astronomy, University of Florida, Bryant Space Science Center, Gainesville, FL 32611-2055, USA}

\begin{abstract}
We report the discovery of rapid variations of a high-velocity \civ \ broad absorption line trough in the quasar SDSS\,J141007.74+541203.3.  This object was 
intensively observed in 2014 as a part of the Sloan Digital Sky Survey Reverberation Mapping Project, during which 32 epochs of spectroscopy were obtained with the Baryon Oscillation Spectroscopic Survey spectrograph.  We observe significant ($>$ 4$\sigma$) variability in the equivalent width of the broad ($\sim$~4000~\kms \ wide) \civ \ trough on rest-frame timescales as short as 1.20 days ($\sim$29 hours), the shortest broad absorption line variability timescale yet reported. 
The equivalent width varied by $\sim$10\% on these short timescales, and by about a factor of two over the duration of the campaign. 
We evaluate several potential causes of the variability, concluding that the most likely cause is a rapid response to changes in the incident ionizing continuum. If the outflow is at a radius where the recombination rate is higher than the ionization rate, the timescale of variability places a lower limit on the density of the absorbing gas of \density. The broad absorption line variability characteristics of this quasar are consistent with those observed in previous studies of quasars, indicating that such short-term variability may in fact be common and thus can be used to learn about outflow characteristics and contributions to quasar/host-galaxy feedback scenarios.
\end{abstract}

\section{INTRODUCTION}
\label{introduction}

Broad absorption lines (BALs) in quasar spectra have observed velocity widths greater than 2000 \kms \ and are thought to be produced by high-velocity outflowing winds launched near the quasar's supermassive black hole (SMBH; e.g., \citealt{Weymann91}). Many theoretical investigations into these winds have supported a model in which the outflows are driven by radiation pressure on lines (e.g., \citealt{Murray95}; \citealt{Proga00}), although magnetohydrodynamical forces may contribute (e.g., \citealt{Konigl94}). These winds are potentially a means by which gas is evacuated from the host galaxy and thus have the potential to affect the growth of both the galaxy and its SMBH (e.g., \citealt{Dimatteo05}; \citealt{Moll07}; \citealt{Springel05b}; \citealt{King10}). BALs are observed in 10--15\% of optically selected quasars (e.g., \citealt{Gibson09b}; \citealt{Allen11});  as such, BALs play an important part in our understanding of both quasars and galaxy/quasar evolution. The detailed structures and locations of these outflows are not well determined, and we must understand their environments to determine if BALs are viable agents of feedback in their host galaxies.

Variability of BAL troughs provides a powerful tool with which to constrain the properties of these outflows. BALs are observed to be variable in time, both in strength (frequently characterized by equivalent width) and in shape. There are many possible causes of the variability, including movement of the absorbing gas (``cloud crossing") or changes in the ionizing radiation received by the gas. Several studies show support for the variability arising due to bulk movement of the outflowing gas (e.g., \citealt{Lundgren07}; \citealt{Gibson08}; \citealt{Hall11}; \citealt{Vivek12}; \citealt{Capellupo13}), while others present evidence consistent with ionization changes or changes in ``shielding gas" as the cause (e.g., the coordinated variability seen in multiple troughs of the same transition by \citealt{Filizak12}, 2013). It is possible that the mechanisms causing variability differ from object to object, yet each mechanism allows one to constrain different properties of the outflow environment. For example, if the variability arises from cloud crossing, one can estimate the radius of the absorbing gas from the ionizing continuum source. If the variability is due to ionization response, one can estimate the density of the outflow. Either of these parameters can help determine whether or not BALs are viable sources of large-scale feedback to their host galaxies. 

The variability timescales probed by previous BAL studies vary widely --- studies have reported BAL-trough variability on timescales of years (e.g., \citealt{Filizak13}) all the way down to timescales of 8--10 days (\citealt{Capellupo13}), with many studies examining timescales in between (e.g., \citealt{Barlow93}; \citealt{Lundgren07}). BAL variability has been seen on all of the timescales probed; however, to date, no studies have reported variability on rest-frame timescales shorter than 8--10 days.  In addition, BALs are generally observed to be less variable on shorter timescales (e.g., \citealt{Gibson10}; \citealt{Filizak13}). Detection of variability on short timescales would yield novel constraints on variability mechanisms and outflow environments.

Recent spectroscopy from the Sloan Digital Sky Survey Reverberation Mapping Project (SDSS-RM; \citealt{Shen15}), whose primary goal is reverberation mapping of the broad emission line regions of quasars, also includes observations of a number of quasars hosting BALs. The cadence of these observations allows investigation of BAL variability in a number of targets down to short rest-frame timescales. In particular, preliminary measurements for one target indicated \civ \ BAL variability on short ($\sim$1 day) rest-frame timescales. This target, the quasar SDSS\,J141007.74+541203.3 (also known as SBS 1408+544A; \citealt{Chavushyan95}), was observed spectroscopically in 1991 by \cite{Stepanian98}, who noted strong absorption in the \civ \ emission-line region. There was also one previous spectrum taken as a part of the SDSS-III Baryon Oscillation Spectroscopic Survey (BOSS;  \citealt{Eisenstein11}; \citealt{Dawson13}) from 2013 May that was released in the SDSS Data Release 12 (DR12; \citealt{Alam15}). Based on the \mgii \ and \ciii \ emission lines from the DR12 spectrum, the redshift of this quasar is $z = \ $2.337 $\pm$ 0.003. It has an apparent $i$-band magnitude $m_{i} = 18.1$ (\citealt{Alam15}) and absolute magnitude $M_i$ = \absmag \ (\citealt{Paris15}).

We here report on the detailed investigation of this target and the implications of these measurements for the environment producing the BAL trough. In Section~\ref{sec:data}, we present the data used in our investigation and its preparation for analysis. We discuss our measurements of the BAL parameters and the search for variability in Section~\ref{sec:measurements} and follow with a review of possible causes of variability and the physical constraints we can derive from our data in Section~\ref{sec:discussion}. We conclude in Section~\ref{sec:conclusions} by summarizing our results and their implications. Throughout this work, we assume a cosmology with $H_{0}~=~70$~km~s$^{-1}$~Mpc$^{-1}$, $\Omega_{\rm M}~=~0.3$, and $\Omega_{\Lambda}~=~0.7.$

\section{DATA AND DATA PREPARATION}
\label{sec:data}

\subsection{BOSS Spectra} 
The spectra utilized in our study were acquired as a part of the SDSS-RM project, which is a dedicated multi-object reverberation mapping campaign performed as a part of SDSS-III BOSS. For a technical overview of the SDSS-RM project, see \cite{Shen15}. In brief, 849 quasars were spectroscopically monitored with the BOSS spectrograph on the SDSS 2.5-meter telescope (\citealt{Smee13}; \citealt{Gunn06}) from 2014~January~1 through 2014~July~3, resulting in 32 epochs of spectroscopic observations over this period. The median spacing between observations is about 4 days --- see Table~2 of \cite{Shen15} for a complete log of these observations. 
The FOV of the BOSS spectrograph is 2.5 degrees in diameter, and its wavelength coverage is 3650--10,400~\AA \ with a spectral resolution of $R \sim 2000$. Because of the need for high-precision relative spectrophotometric calibrations for the SDSS-RM project, the data were processed using an improved flux-calibration method that uses extra simultaneous standard star observations (see \S 3.3 and 3.4 of \citealt{Shen15}). The first SDSS-RM BOSS spectrum was obtained on MJD\,56660.209 --- we define this as the ``beginning" of our campaign, and all light curves presented as a part of our analysis (unless explicitly stated otherwise) show the rest-frame time between each epoch and MJD\,56660.209 (2~January~2014). 
\begin{figure*}
\begin{center}
\includegraphics[scale = 0.325, angle = -90, trim = 0 0 35 10, clip]{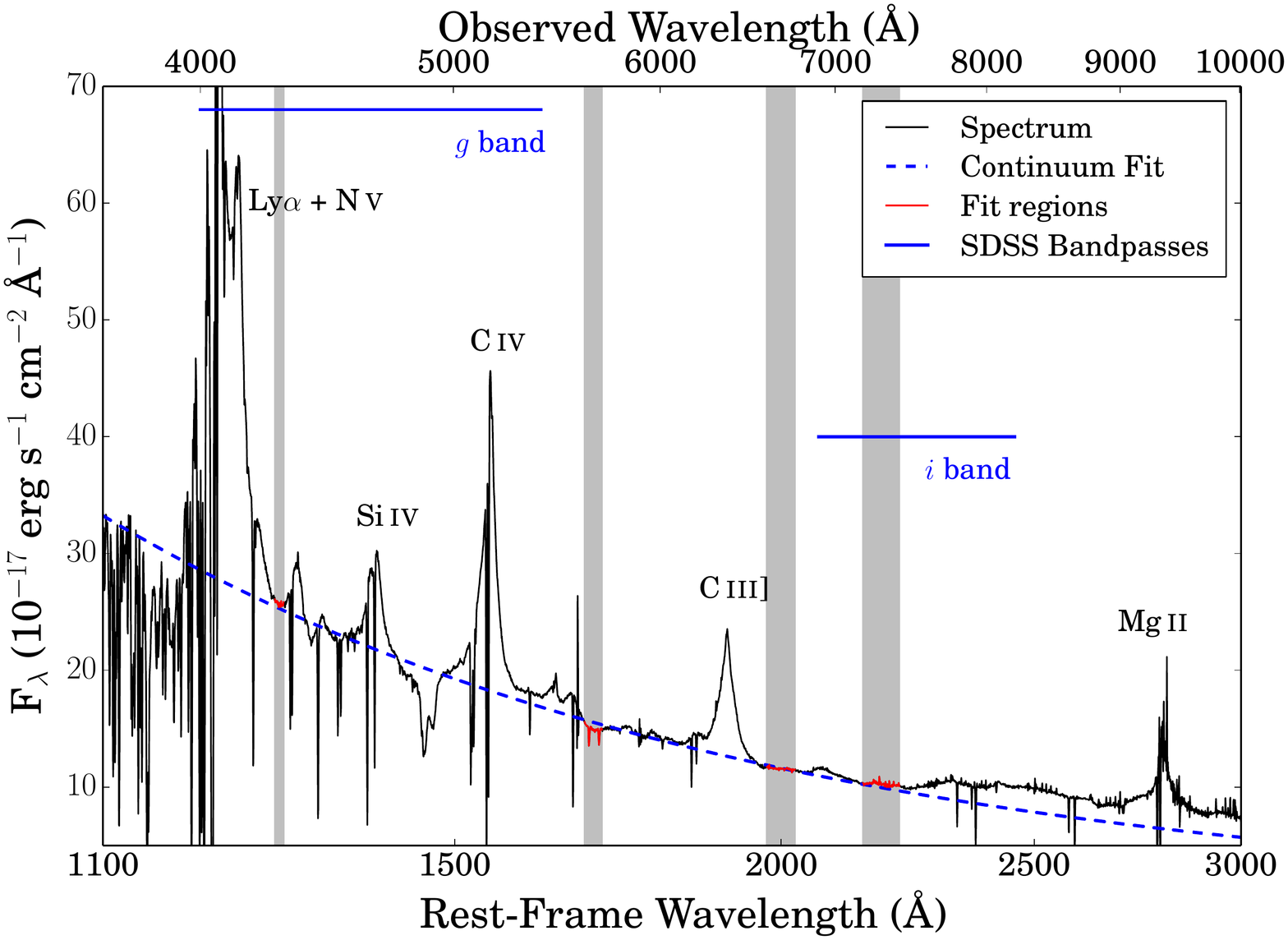}
\includegraphics[scale = 0.325, angle = -90, trim = 0 0 35 10, clip]{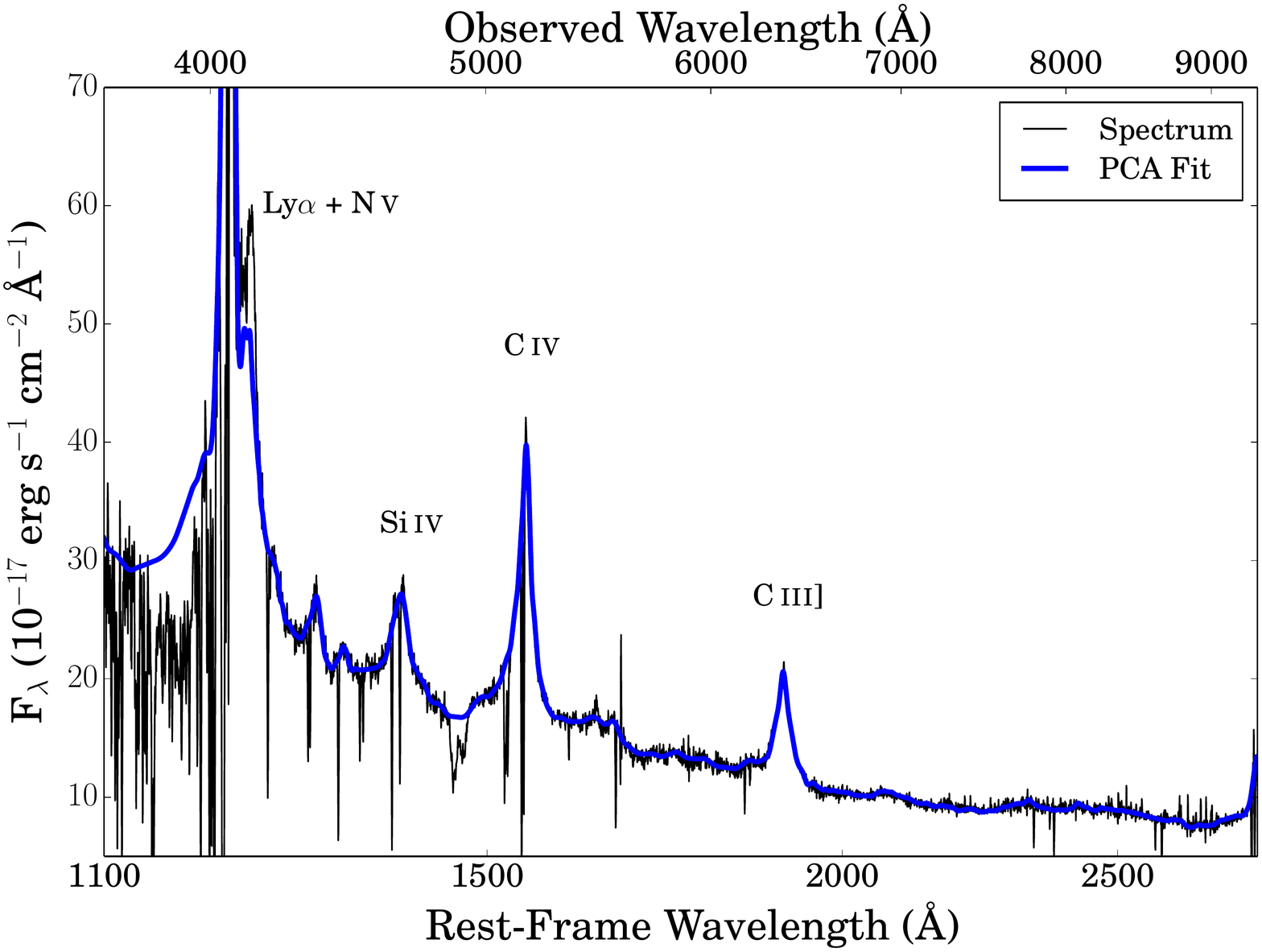}
\caption{Left panel: The mean spectrum (black solid line) from all 31 epochs used in this analysis with its continuum fit (blue dashed line). The line-free regions chosen to include in the continuum fit are shaded in gray and also highlighted in red.  The approximate wavelength coverage of the SDSS $g$- and $i$-band photometric filters are shown as solid blue horizontal lines. Right panel: The SDSS-RM spectrum (black) from MJD 56683 and its accompanying PCA fit (blue) from \cite{Paris15}. We use the PCA fit to remove the \civ \ emission-line signatures, and investigate the lower-velocity \civ \ absorption features.}
\label{fig:confit}
\end{center}
\end{figure*}

The processed BOSS spectra from this campaign were first corrected for Galactic extinction using a $R_{\rm V}$ = 3.1 Milky Way extinction model (\citealt{Cardelli89}) and $A_{\rm V}$ values from \cite{Schlegel98}.  Before beginning analysis, we converted the observed wavelengths of the spectra to the rest frame. All 32 epochs of data were visually inspected for quality; the median signal-to-noise ratio (S/N) of the spectra was $\sim$30 per pixel, with most spectra falling between 20--40. We excluded one spectrum from our analysis due to low S/N ($<$ 10), leaving 31 SDSS-RM spectral epochs for our analysis and the earlier observation from the DR12 data release. Figure~\ref{fig:confit} shows the mean spectrum of our quasar during the SDSS-RM campaign (RMID 613 in Table 1 of \citealt{Shen15}). 
Visual inspection of the spectrum reveals a BAL quasar with low reddening and a prominent \civ \ BAL trough that is at sufficiently high velocity to be detached from the \civ \ emission line. There are no BAL features in lower-ionization transitions like \mgii \ or \aliii, making this target a high-ionization BAL quasar. There are also two narrower \civ \ absorption features at lower outflow velocities superimposed on the \civ \ emission line, as well as \nv \ absorption redward of the Lyman-$\alpha$ emission line, but this region is heavily contaminated by intervening Lyman-$\alpha$ absorption and thus its analysis is outside the scope of this work. There is a hint of broad \siiv \ absorption at a similar velocity as the \civ \ BAL, but it is heavily contaminated by C\,{\sc ii} and O\,{\sc i}/Si\,{\sc ii} features and does not ever reach a depth below 90\% of the continuum level. 

\begin{figure}
\begin{center}
\includegraphics[scale = 0.4, angle = -90, trim = 0 0 195 260, clip]{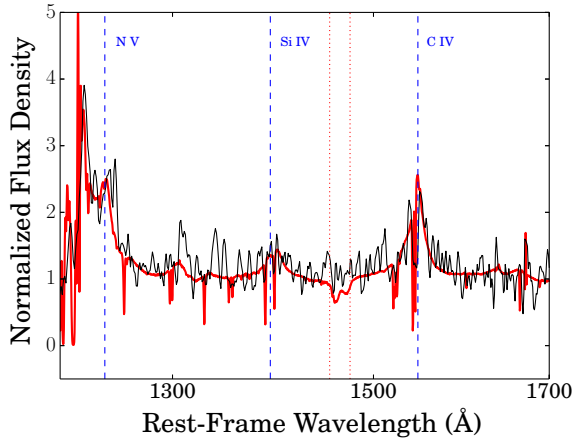}
\caption{The 1991 spectrum from \cite{Chavushyan95}, smoothed by 5 pixels and normalized by the mean flux density (black solid line) and our mean continuum-normalized spectrum from the SDSS-RM campaign (red solid line). The blue dashed lines show the rest-frame wavelengths of \nv, \siiv, and \civ \ for reference. The red dotted lines show the wavelength region spanned by the \civ \ BAL trough during the SDSS-RM campaign. }
\label{fig:oldspec}
\end{center}
\end{figure}

We also inspected the spectrum taken on 1991 October 4 by  \cite{Chavushyan95} for comparison (Figure~\ref{fig:oldspec}). The \civ \ BAL trough does not appear to be present in the 1991 spectrum, though the S/N is too low to determine definitively whether or not the BAL trough is present; however, the depth measured within the trough region in the 1991 spectrum is consistent with there being no trough flux below 90\% of the continuum level to $\sim$3$\sigma$ significance. The lowest-velocity, narrower \civ \ absorption feature close to the \civ \ line center is clearly present in the 1991 spectrum, though we cannot tell if the nearby higher-velocity narrow absorption feature superimposed on the \civ \ emission line is present due to low S/N. 

\subsection{Continuum Fits and Normalization} 
\label{sec:confits}
To disentangle the BAL variability from the overall quasar variability, we fit a continuum to each individual spectrum. Following \S 3.1 of \cite{Filizak12}, we first attempted to fit a reddened power law model to the continuum emission using the SMC-like reddening model from \cite{Pei92} --- however, the results were consistent with no extinction, so a simple, non-reddened power law yielded suitable fits for this particular target. We used a nonlinear least-squares algorithm and fit to four regions within the spectra that were relatively free of emission and absorption features. We chose these relatively line-free regions, shown in Figure~\ref{fig:confit}, by visually inspecting the mean spectrum and comparing it with the SDSS composite spectrum of \cite{Vandenberk01}. We used the same line-free regions for each epoch. Because some of the line-free regions contained more pixels than others, we weighted each pixel such that each individual region contributed equally in the continuum fit. This ensures that none of the line-free regions holds more weight in the fit just because it spans a larger wavelength range (and thus contains more pixels). 

\begin{figure*}
\begin{center}
\includegraphics[scale = 0.25, angle = -90, trim = 0 0 100 0, clip]{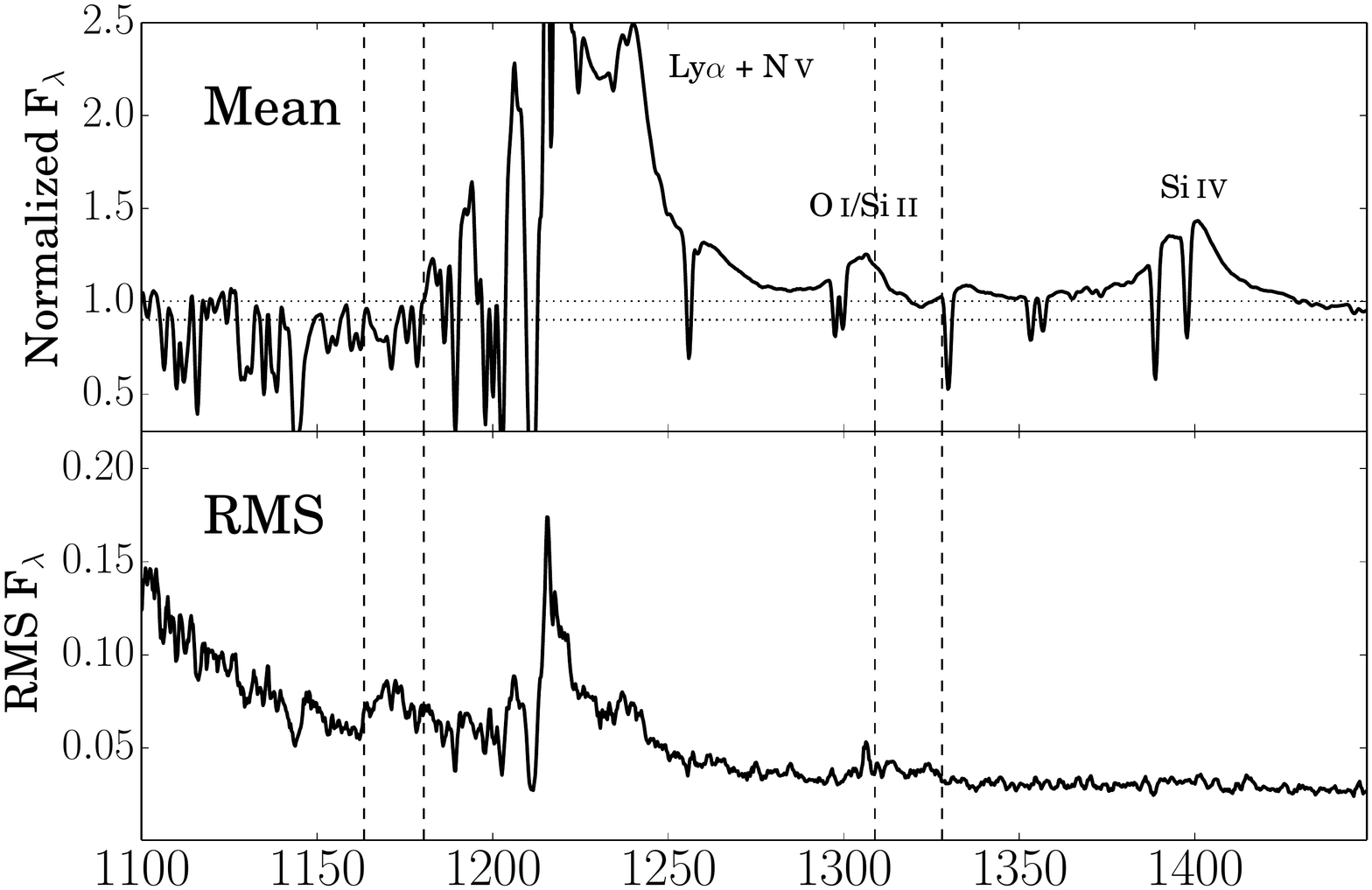}
\includegraphics[scale = 0.25, angle = -90, trim = 0 0 90 0, clip]{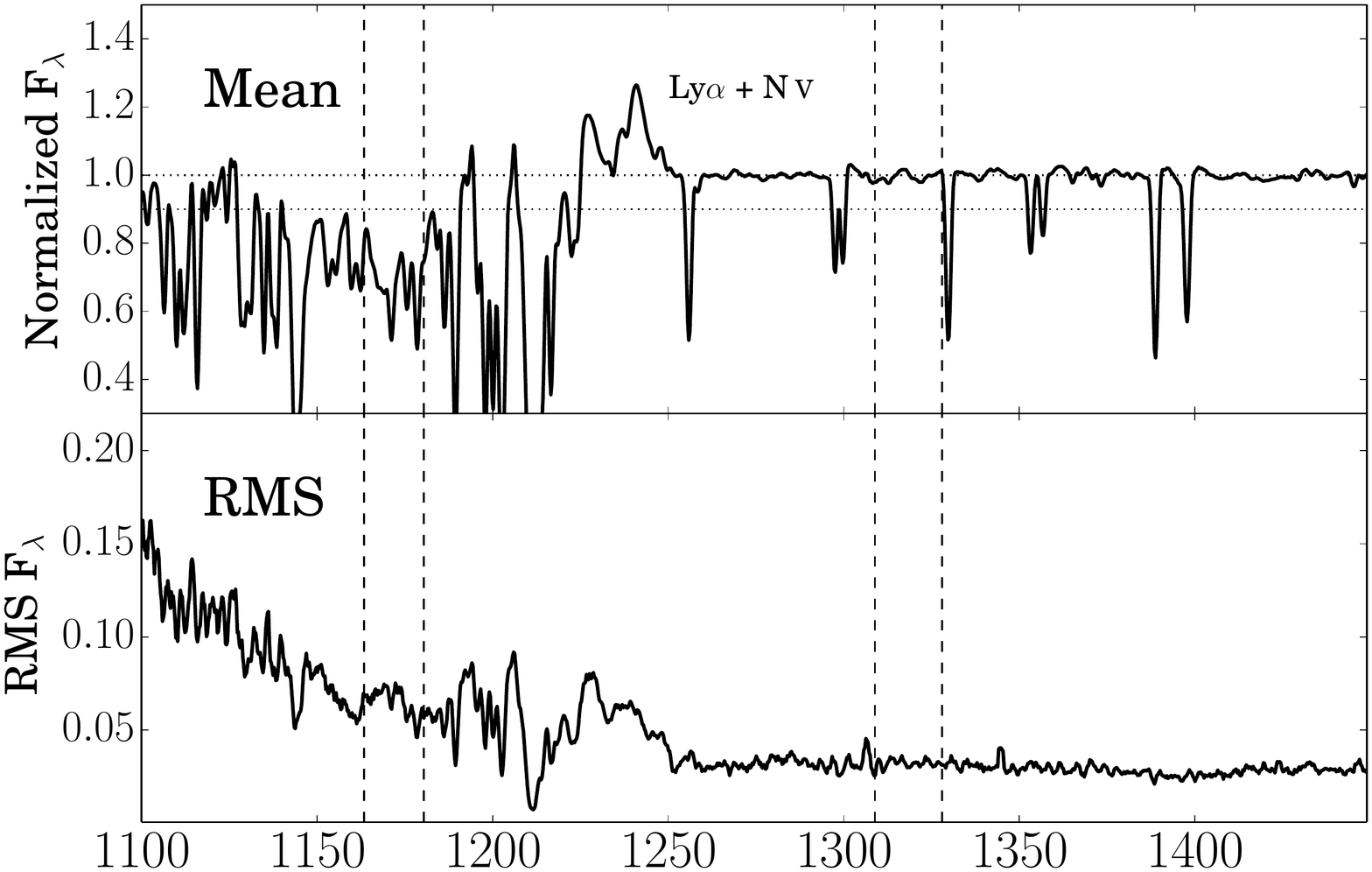}
\includegraphics[scale = 0.25, angle = -90, trim = 0 0 100 0, clip]{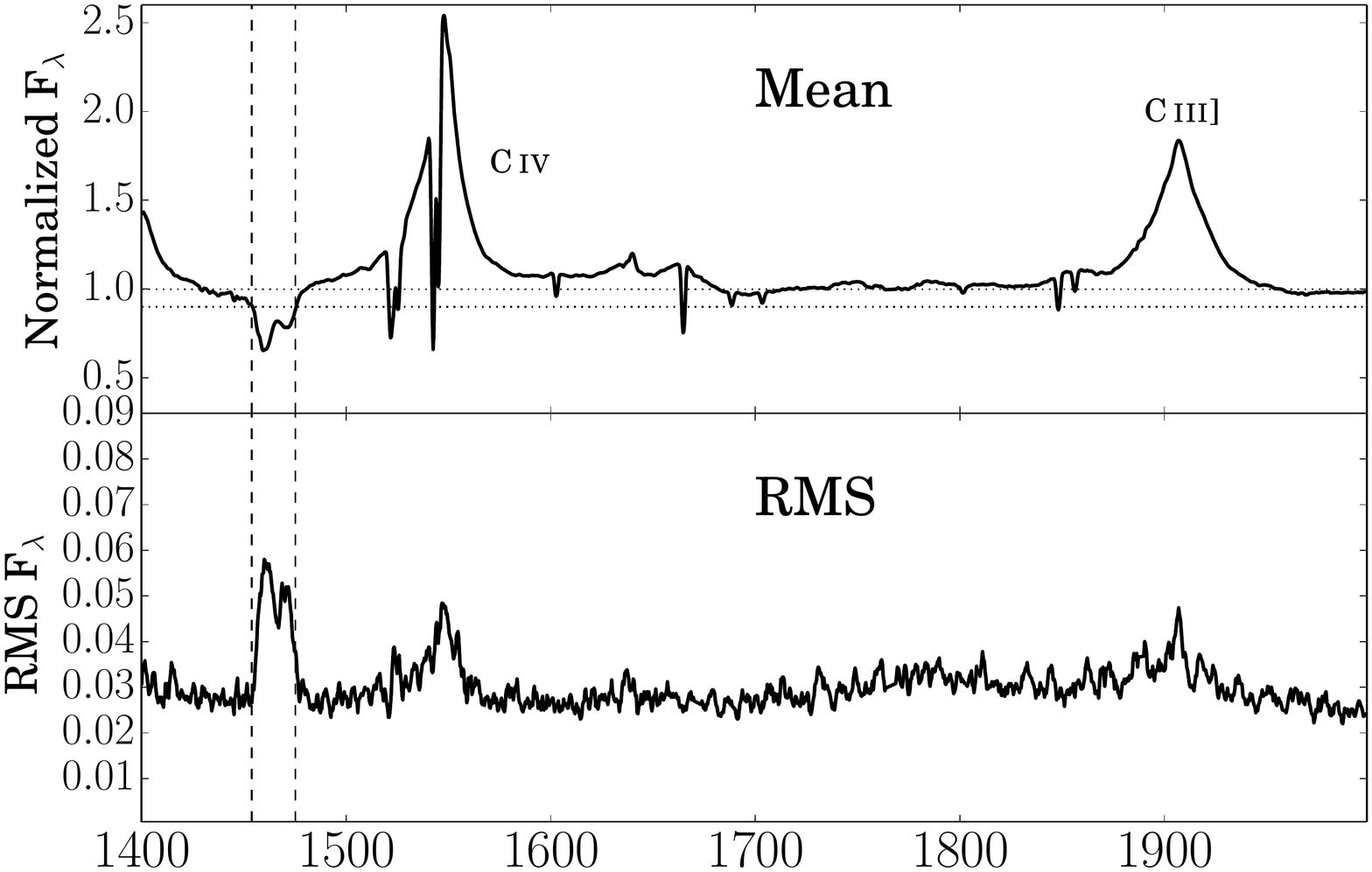}
\includegraphics[scale = 0.25, angle = -90, trim = 0 0 90 0, clip]{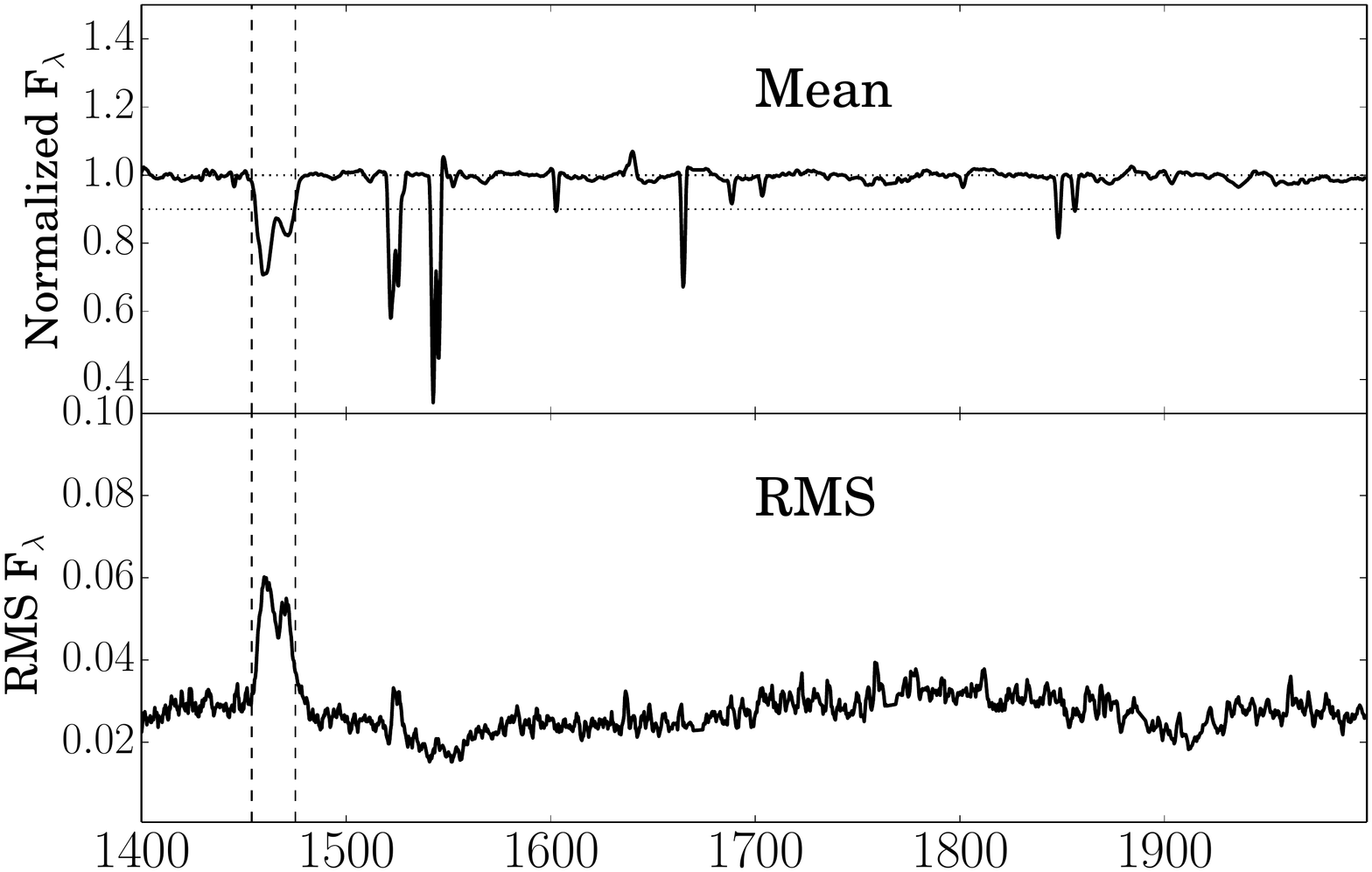}
\includegraphics[scale = 0.25, angle = -90, trim = 0 0 60 0, clip]{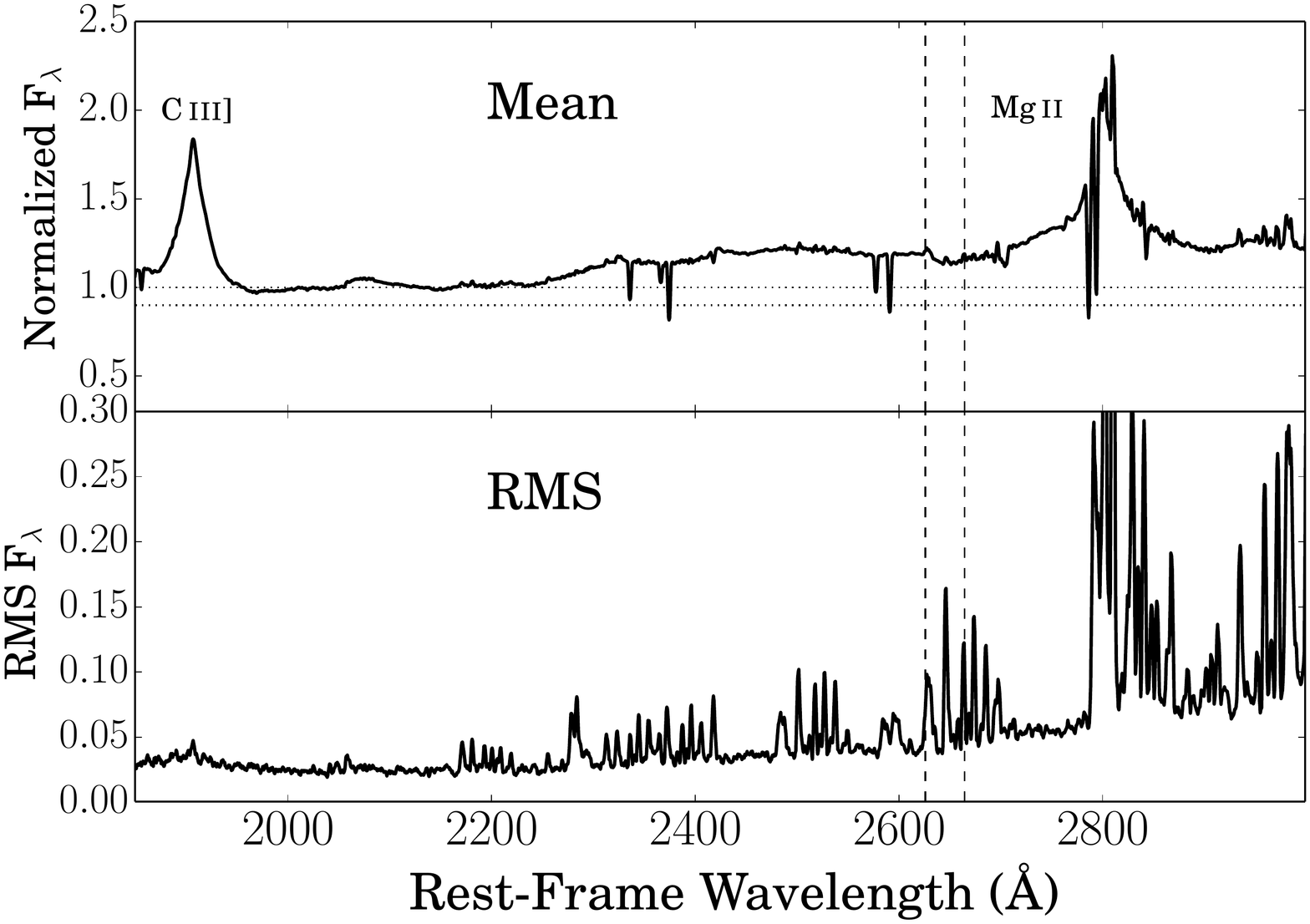}
\includegraphics[scale = 0.25, angle = -90, trim = 0 0 70 0, clip]{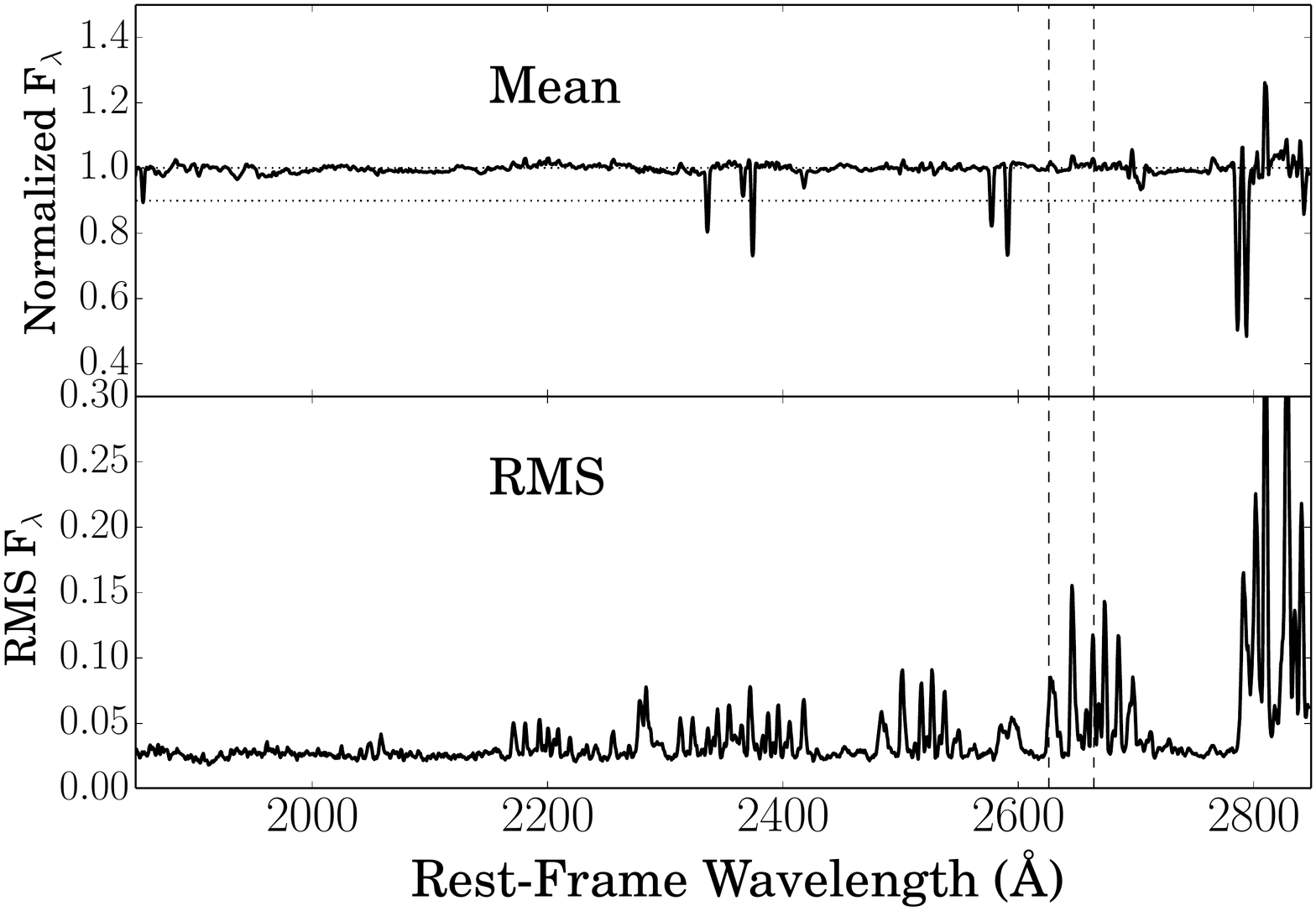}
\caption{Mean and RMS residual spectra for the continuum-normalized (left panels) and of the PCA-normalized (right panels) cases. The PCA fit generally results in similar mean and residual spectra compared to our continuum fit, but also allows one to isolate and characterize the low-velocity \civ \ absorption near the broad emission. Each panel shows a different wavelength region along the spectrum. The top subpanel for each region displays the mean continuum-normalized spectrum over the 31 epochs, and the bottom panel shows the RMS residual spectrum. Dotted horizontal lines across the mean spectrum show a normalized flux density of 1.0 and 0.9 to help the eye in identifying absorption features. Vertical dashed lines indicate the wavelength regions spanned by the \civ \ BAL (middle panels) and also in which absorption associated with the \civ \ BAL would be expected to appear for \nv \ and \siiv \ (top panels) and \mgii \ (bottom panels). }
\label{fig:confit_mean_rms}
\end{center}
\end{figure*}

To characterize the uncertainties in the continuum fits, we account for two different sources of error. The first component accounts for the uncertainty caused by noise in the spectra. To evaluate these uncertainties, we used ``flux randomization'' Monte Carlo iterations as described by \cite{Peterson98b}, where the flux in each pixel of the spectrum is altered by a random Gaussian deviate based on the uncertainty assigned to that particular pixel. The continuum is then fit to the new spectrum, and the process is repeated 100 times. We adopt the standard deviation of these 100 iterations as the first component of the uncertainties in the continuum fit. 

The second component to the continuum uncertainties considers the small changes in the fit resulting from the decisions we made when fitting the continuum. For example, we choose which particular line-free regions of the spectrum are used in the fit, and whether or not each pixel is given equal weight. To evaluate the uncertainties introduced by these decisions, we created two more continuum fits for each spectrum. First, we fit the continuum in each epoch using two additional line-free regions, but still using the same weighting scheme as described above, where each line-free region is given equal weight.  Second, we fit a continuum using our original line-free regions while weighting each pixel equally. We then adopted the average deviation of these two continuum fits from the original continuum fit as the second component to the continuum uncertainty.

The two uncertainty components are added in quadrature and adopted as our formal uncertainties in each continuum fit. Although a continuum was fit to each epoch individually, Figure~\ref{fig:confit} shows the fit to the mean spectrum as an example of how well the fits performed. Each individual epoch was divided by its continuum fit to obtain a normalized spectrum, and both the spectral and continuum fit uncertainties were propagated to determine the final uncertainty on the normalized spectrum. Hereafter, we will refer to these normalized spectra as the ``continuum-normalized" spectra.  In all cases, the uncertainties in the continuum-normalized spectra are dominated by the spectral uncertainties rather than the formal uncertainties in the continuum model. 
We created a mean and root mean square (RMS) residual spectrum from the continuum-normalized spectra; the RMS residual spectrum is obtained by taking the standard deviation about the mean spectrum, the result of which shows the variable components (Figure~\ref{fig:confit_mean_rms}). The base level of the RMS flux (on the order of $\sim0.2$) in all panels is above zero due to noise in the spectra, which adds a constant to the entire RMS spectrum. The RMS spectrum contained a few sharp features arising from poor/inconsistent sky subtraction; these small segments of the spectra were masked to avoid confusion, as they did not lie near any of the emission or absorption features of interest in this study (all fell between 1600--1900~\AA \ in the rest frame). The region of the spectrum redward of $\sim$2200~\AA \ suffered from telluric contamination; these features were too widespread to mask, so they are shown in full in Figure~\ref{fig:confit_mean_rms}. 

\subsection{Principal Component Analysis Fits} 
As a part of the SDSS DR12 quasar catalog production (\citealt{Paris15}), a principal component analysis (PCA) was done on each epoch of BOSS spectroscopy. This procedure was also used for the DR9 and DR10 quasar catalogs \citep{Paris12, Paris14}; see these works for a description of this technique. PCA  fits both the emission-line and continuum features in each spectrum using a reference sample of quasars for which the relevant features are observed. An example of a PCA fit to one of our spectra is shown in Figure~\ref{fig:confit}. In this target, two lower-velocity and narrower \civ \ absorption features are seen superimposed on the \civ \ emission line, and the \civ \ emission-line flux in this low-velocity region is well-fit by the PCA approach. Thus, we consider the PCA fits in our analysis to help eliminate contributions from the \civ \ emission line, particularly relevant for measurements of the lower-velocity absorption features. As with our continuum fits, we divide the original spectra by their PCA fits to obtain the ``PCA-normalized" spectra at each epoch for use when investigating the lower-velocity \civ \ absorption features. Figure~\ref{fig:confit_mean_rms} shows the mean and RMS residual spectrum of the PCA-normalized spectra. All of the emission features are well-removed from the mean and RMS residual spectra by the normalization except for the Lyman-$\alpha$ region, which is blended with \nv \ and has significant Lyman-$\alpha$ absorption on top of the emission line. The poor fit is due to the  sample size used in the PCA in this region --- below rest-frame 1280 \AA , the sample used to perform PCA is small and makes use of hand-fitted continua, so we are able to reproduce
fewer features in this region (e.g., \citealt{Paris12}). As with the continuum-normalized spectra, we mask a few regions with significant sky features in the RMS residual spectrum to avoid confusion. 

\subsection{Photometric Observations}
\label{sec:photobs}
As a part of the SDSS-RM program, photometric observations of our target were also obtained with the 2.3m Steward Observatory Bok telescope and the 3.6m Canada-France-Hawaii Telescope (CFHT). The observations are described in \S 3.5 of \cite{Shen15}. The Bok/90Prime instrument has a 1 deg$^{2}$ field of view and a plate scale of $0\farcs45$ pixel$^{-1}$. About 60 nights were allocated to the SDSS-RM program from January-July 2014, yielding 31 epochs in the $g$ band and 27 in the $i$ band by the end of the program. The CFHT observations were taken with MegaCam, which also has a 1 deg$^{2}$ field of view, with a pixel size of $0\farcs187$. Over the course of the campaign, 26 epochs were observed in the $g$ band and 20 observed in the $i$ band with the CFHT (see \citealt{Fukugita96} for a description of the $ugriz$ filters). 

\begin{figure}
\begin{center}
\includegraphics[scale = 0.35, angle = -90, trim = 0 0 310 130, clip]{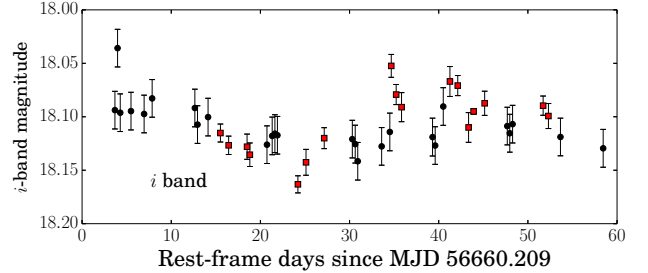}
\caption{Photometric $i$-band continuum light curves from the CHFT (black filled circles) and Bok (red squares) telescopes. The BAL quasar varies by only $\sim$10\% over this time period.}
\label{fig:phot_lcs}
\end{center}
\end{figure}

In this target, the $g$ filter covers both Lyman-$\alpha$/\nv \  and \civ \ (see Figure~\ref{fig:confit}), both of which are observed to vary during the campaign, particularly the \mbox{Lyman-$\alpha$/\nv} \ region (Figure~\ref{fig:confit_mean_rms}). Thus, we do not consider the $g$-band photometric light curve to be a good proxy for the continuum variability. However, the $i$-band photometric light curve covers a relatively line-free region of the spectrum for this object (Figure~\ref{fig:confit}) so we include this light curve in our investigations (Figure~\ref{fig:phot_lcs}). Aperture photometry was performed on all images using SExtractor (\citealt{Bertin96}) and a preliminary flux calibration was performed. A zero-point shift was applied to place the Bok photometry onto the CFHT magnitude scale, which in turn is tied to the photometry of SDSS objects in our field as a step in the MegaPipe reduction pipeline (\citealt{Gwyn08}). Several epochs were identified as outliers for all targets observed and thus categorized as bad data and removed from the light curves; we had good measurements for 43 total epochs.  The quasar did not exhibit very much photometric variability during the campaign; the flux changed by a factor of only about 10--15\% during this period. Measurements of the flux densities within the line-free regions of the spectra similarly yield light curves consistent with little variability.  

\section{VARIABILITY MEASUREMENTS}
\label{sec:measurements} 

\subsection{Searching for Variability} 
\begin{figure}
\begin{center}
\includegraphics[scale = 0.3, angle = -90, trim = 0 0 80 20, clip]{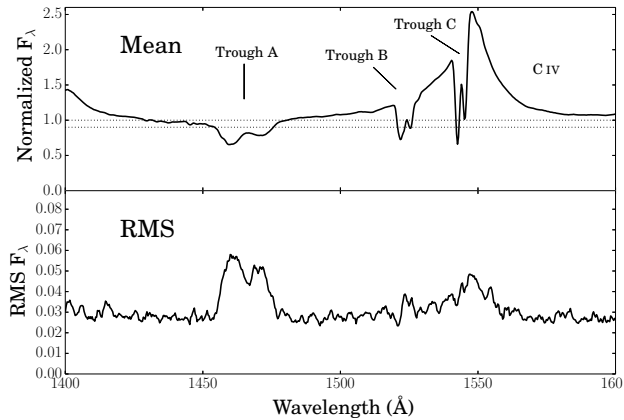}
\caption{Mean and RMS residual spectra for the continuum-normalized case, expanded to emphasize the \civ \ region of the spectrum. Dotted horizontal lines across the mean spectrum show a normalized flux density of 1.0 and 0.9 to help the eye in identifying absorption features.}
\label{fig:mean_zoom}
\end{center}
\end{figure}

Inspection of the \civ \ emission-line region in the mean and RMS residual of both the continuum-normalized and PCA-normalized spectra (Figures \ref{fig:confit_mean_rms} and \ref{fig:mean_zoom}) reveals significant variability in the high-velocity  \civ \ absorption trough centered near $-16700 \ $km s$^{-1}$ (hereafter referred to as Trough A;  see Figure~\ref{fig:mean_zoom}).  Figures \ref{fig:confit_mean_rms} and \ref{fig:mean_zoom} also show some variability in the \civ \ emission line, which makes it more difficult to determine whether there is any variability in the lower-velocity \civ \ absorption troughs using the continuum-normalized spectra. However, the RMS spectrum of the PCA-normalized spectra largely removes the variability signatures from the emission-line features (Figure~\ref{fig:confit_mean_rms}), allowing us to investigate variability in the lower-velocity absorption features. There is detectable low-amplitude variability in the associated medium-velocity \civ \ trough (hereafter Trough B; centered near $-4800$ km s$^{-1}$), but none is detected in the lowest-velocity associated narrow \civ \ absorption trough (hereafter Trough C; centered near $-1000$ km s$^{-1}$) superimposed on the \civ \ emission line. The RMS residual signal in the Trough A region is nearly identical to that in the continuum-normalized RMS spectrum, which is expected given that this trough is detached from the \civ \ emission line and thus we expect little contamination from the emission-line variability.

Examining the RMS residual spectra derived from both the continuum-normalized spectra and the PCA-normalized spectra at various other wavelength regimes allows a search for associated (or non-associated) absorption and emission-line variability across the spectrum. There are signs of variable absorption in the \nv, \aliii, and \siiv \ regions (as evidenced by a slight increases in the RMS residual flux in these regions), but these signatures are not significant enough to disentangle them properly from the noise, and as discussed in Section \ref{sec:confits}, none hosts a BAL trough of $>$10\% fractional depth. We hereafter focus our attention on the \civ \ region of the spectrum shown in Figure~\ref{fig:mean_zoom}. 

\subsection{Measurements}
\label{sec:balmeasurements}

\subsubsection{BAL-Trough Properties} 
\label{sec:baltroughproperties}
To determine the minimum and maximum velocities of Trough A, we smoothed the mean continuum-normalized spectrum using a boxcar smoothing algorithm over five pixels. As is standard in BAL studies, we then determined the velocities between which the flux in Trough A remained below 90\% of the continuum level (i.e., where the normalized flux density is $<$ 0.9)  and obtained \mbox{$v_{\rm min}$~=~$-14470$ \kms} and \mbox{$v_{\rm max}$~=~$-18810$ km s$^{-1}$,} which yields a trough velocity width of 4340 \kms. The minimum and maximum velocities of the trough in each epoch were measured separately, and were consistent with those measured for the mean spectrum to within the uncertainties (the range of measured values was within $\sim$200 km s$^{-1}$ of the mean values). We applied the velocity limits found in the mean spectrum to measure the rest-frame equivalent width (EW), mean fractional depth, and absorbed-flux-weighted centroid velocity ($v_{\rm cent}$) of Trough A at each epoch using the non-smoothed continuum-normalized spectra (Table~\ref{tbl:bal_params}). Uncertainties in each parameter were calculated using flux-randomization Monte Carlo iterations as described in Section \ref{sec:confits}. We adopt the standard deviation of these iterations as our measurement uncertainties and tabulate them in Table \ref{tbl:bal_params}. 

\begin{deluxetable}{lcccc} 
\tablewidth{0pt} 
\tablecaption{BAL Trough Parameters\tablenotemark{a}} 
\tablehead{ 
\colhead{MJD} & 
\colhead{$\Delta t_{\rm rest}$\tablenotemark{b} } & 
\colhead{EW } & 
\colhead{Mean} & 
\colhead{$v_{\rm cent}$} \\ 
\colhead{} & 
\colhead{(days)} & 
\colhead{(\AA) } & 
\colhead{Depth} & 
\colhead{(km s$^{-1}$)} 
} 
\startdata 
56426.00 & 0.00 &  4.37 $\pm$ 0.08 & $0.203$ $\pm$ 0.004 &  $-16799 \pm$  25 \\ 
56660.21 & 70.19 &  5.07 $\pm$ 0.08 & $0.235$ $\pm$ 0.004 &  $-16768 \pm$  20 \\ 
56664.51 & 1.29 &  4.80 $\pm$ 0.11 & $0.223$ $\pm$ 0.005 &  $-16768 \pm$  30 \\ 
56669.50 & 1.49 &  4.77 $\pm$ 0.16 & $0.221$ $\pm$ 0.007 &  $-16698 \pm$  41 \\ 
56683.48 & 4.19 &  5.06 $\pm$ 0.08 & $0.234$ $\pm$ 0.004 &  $-16695 \pm$  17 \\ 
56686.47 & 0.90 &  5.37 $\pm$ 0.08 & $0.249$ $\pm$ 0.004 &  $-16707 \pm$  20 \\ 
56696.78 & 3.09 &  5.78 $\pm$ 0.08 & $0.268$ $\pm$ 0.004 &  $-16707 \pm$  17 \\ 
56715.39 & 5.58 &  6.24 $\pm$ 0.07 & $0.289$ $\pm$ 0.003 &  $-16706 \pm$  17 \\ 
56717.33 & 0.58 &  5.81 $\pm$ 0.08 & $0.269$ $\pm$ 0.003 &  $-16711 \pm$  15 \\ 
56720.45 & 0.93 &  6.02 $\pm$ 0.09 & $0.279$ $\pm$ 0.004 &  $-16724 \pm$  17 \\ 
56722.39 & 0.58 &  5.95 $\pm$ 0.06 & $0.276$ $\pm$ 0.003 &  $-16737 \pm$  11 \\ 
56726.46 & 1.22 &  5.81 $\pm$ 0.09 & $0.269$ $\pm$ 0.004 &  $-16727 \pm$  16 \\ 
56739.41 & 3.88 &  5.64 $\pm$ 0.08 & $0.261$ $\pm$ 0.004 &  $-16688 \pm$  19 \\ 
56745.28 & 1.76 &  5.28 $\pm$ 0.08 & $0.245$ $\pm$ 0.004 &  $-16687 \pm$  19 \\ 
56747.42 & 0.64 &  5.09 $\pm$ 0.08 & $0.236$ $\pm$ 0.004 &  $-16743 \pm$  25 \\ 
56749.37 & 0.59 &  5.40 $\pm$ 0.09 & $0.250$ $\pm$ 0.004 &  $-16721 \pm$  19 \\ 
56751.34 & 0.59 &  5.49 $\pm$ 0.08 & $0.254$ $\pm$ 0.003 &  $-16735 \pm$  18 \\ 
56755.34 & 1.20 &  4.95 $\pm$ 0.09 & $0.229$ $\pm$ 0.004 &  $-16741 \pm$  21 \\ 
56768.23 & 3.86 &  5.22 $\pm$ 0.07 & $0.242$ $\pm$ 0.003 &  $-16756 \pm$  17 \\ 
56772.23 & 1.20 &  4.79 $\pm$ 0.07 & $0.222$ $\pm$ 0.003 &  $-16729 \pm$  19 \\ 
56780.23 & 2.40 &  4.41 $\pm$ 0.09 & $0.205$ $\pm$ 0.004 &  $-16746 \pm$  22 \\ 
56782.25 & 0.60 &  4.32 $\pm$ 0.08 & $0.200$ $\pm$ 0.004 &  $-16768 \pm$  22 \\ 
56783.25 & 0.30 &  4.51 $\pm$ 0.08 & $0.209$ $\pm$ 0.004 &  $-16755 \pm$  21 \\ 
56795.18 & 3.57 &  4.01 $\pm$ 0.09 & $0.186$ $\pm$ 0.004 &  $-16776 \pm$  25 \\ 
56799.21 & 1.21 &  4.07 $\pm$ 0.08 & $0.188$ $\pm$ 0.004 &  $-16783 \pm$  22 \\ 
56804.19 & 1.49 &  3.90 $\pm$ 0.08 & $0.181$ $\pm$ 0.004 &  $-16774 \pm$  25 \\ 
56808.26 & 1.22 &  4.08 $\pm$ 0.08 & $0.189$ $\pm$ 0.003 &  $-16739 \pm$  25 \\ 
56813.23 & 1.49 &  3.77 $\pm$ 0.08 & $0.175$ $\pm$ 0.004 &  $-16750 \pm$  27 \\ 
56825.19 & 3.58 &  3.68 $\pm$ 0.07 & $0.171$ $\pm$ 0.003 &  $-16790 \pm$  29 \\ 
56829.21 & 1.21 &  3.34 $\pm$ 0.08 & $0.155$ $\pm$ 0.004 &  $-16866 \pm$  33 \\ 
56833.21 & 1.20 &  3.41 $\pm$ 0.08 & $0.158$ $\pm$ 0.004 &  $-16782 \pm$  32 \\ 
56837.19 & 1.19 &  3.65 $\pm$ 0.07 & $0.169$ $\pm$ 0.003 &  $-16796 \pm$  24 \\ 
mean\tablenotemark{c} &  \nodata  &  4.83 $\pm$  0.01 & $0.224$ $\pm$ 0.001 &  $-16739 \pm$   4 
\enddata 
\tablenotetext{a}{All measurements are made with the continuum-normalized spectra.} 
\tablenotetext{b}{All $\Delta t$ measurements are made with respect to the previous epoch in the rest-frame.} 
\tablenotetext{c}{Measurements made from the mean spectrum created from the 31 SDSS-RM spectra --- excluding the early spectrum from DR12.} 
\label{tbl:bal_params} 
\end{deluxetable}  

\begin{figure}
\begin{center}
\includegraphics[scale =  0.42, angle = 0, trim = 0 0 0 250, clip]{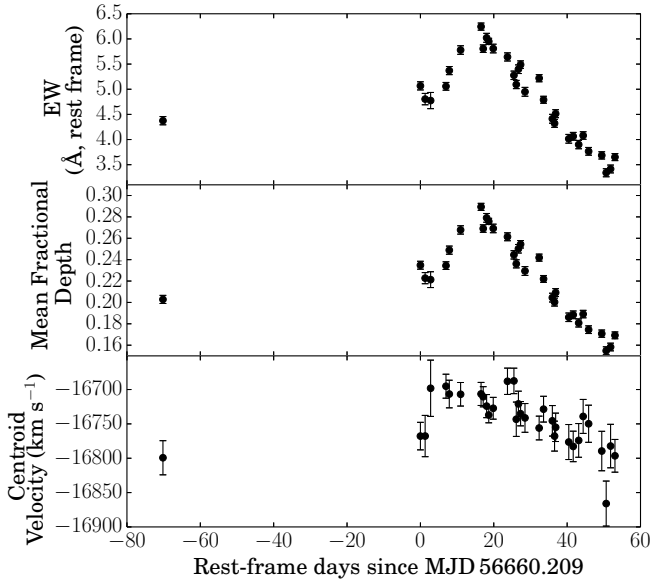}
\caption{Equivalent width (top panel), mean fractional depth (middle panel) and centroid velocity (bottom panel) of the high-velocity  \civ \ BAL trough (Trough A; centered near $-$16700 \kms) as a function of days since the RM campaign began, in the rest frame. The point at far left is the additional observation of our target taken by SDSS, about 70 rest-frame days before the SDSS-RM project began. The apparent decline in velocity centroid, on the order of $\sim$100 \kms,  is of much smaller amplitude than the changes seen in the EW and depth, and corresponds to less than two pixels in the spectral direction.}
\label{fig:lightcurves}
\end{center}
\end{figure}

Figure~\ref{fig:lightcurves} shows the EW, mean fractional depth, and $v_{\rm cent}$  measurements of Trough A as a function of rest-frame time from the beginning of the RM campaign. Both the trough EW and depth increase from $\Delta t$ = 0 to $\sim$15 days by about 23\%, followed by a steady decline over the rest of the campaign by a factor of $\sim$2. The similarity of the EW and mean fractional depth light curves indicates that the trough profile does not change significantly over the period of the campaign;  we investigate this further below. The centroid velocity decreases slightly over the campaign, but with a small amplitude of $\sim$ 100 \kms  \ that corresponds to less than two pixels in the spectral direction. 

\begin{figure}
\begin{center}
\includegraphics[scale = 0.46, angle = -90, trim = 0 0 85 250, clip]{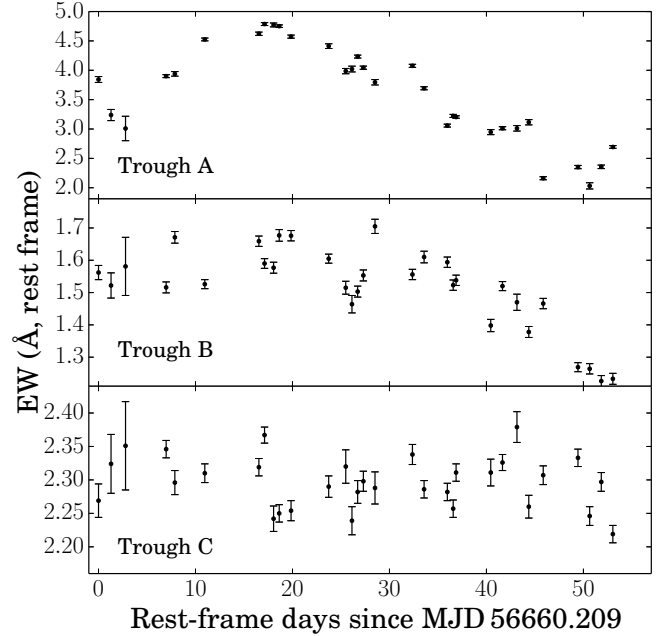}
\caption{Equivalent widths of Trough A (centered at $\sim-$16700 \kms), Trough B (centered at $\sim-4800$ \kms), and Trough C (centered at $\sim-1025$ \kms) absorption components as a function of rest-frame days since the RM campaign began. These measurements were made using the PCA-normalized spectra rather than the continuum-normalized spectra, as Troughs B and C lie on top of the \civ \ emission line.}
\label{fig:pca_ewlightcurves}
\end{center}
\end{figure}

We also measure the EW in all three \civ \ absorption troughs (A, B, and C) using the PCA-normalized spectra, as the latter two cannot be examined with the continuum-normalized spectra due to their positions on top of the \civ \ emission line. We examine the resulting light curves (Figure~\ref{fig:pca_ewlightcurves}): The widths of Trough B and Trough C are $\sim$1311 \kms \ and $\sim$1380 \kms, respectively, classifying them as mini-BALs. For our purposes, we do not need to consider each component of the \civ \ doublet separately in these troughs.  Nonetheless, we note that Trough B appears to be a line-locked system of two \civ \ absorbers, with the weaker one closer to the systemic redshift. The light curve for Trough A created from the PCA-normalized spectra is very similar to that obtained using the continuum-normalized spectra; there is again a clear increase, followed by a long-term decrease in EW. However, the PCA fits produce more scatter in the EW light curve of Trough A than the continuum fits, likely due to uncertainty in determining the boundaries of the BAL trough for exclusion during the PCA fitting. Thus, we rely on our continuum-normalized spectra for our quantitative assessment of Trough A.  Our conclusions below about the BAL variability still hold when considering the PCA-normalized spectra --- however, we opt for the conservative route of using the continuum-normalized spectra for our quantitative analysis. However, when comparing the behavior of Trough A with Troughs B and C, we use the PCA fits for all three troughs to avoid introducing possible systematic uncertainties between the continuum-normalized and PCA-normalized spectra. 

\begin{figure}
\begin{center}
\includegraphics[scale = 0.43, angle = -90, trim =0 0 200 200, clip]{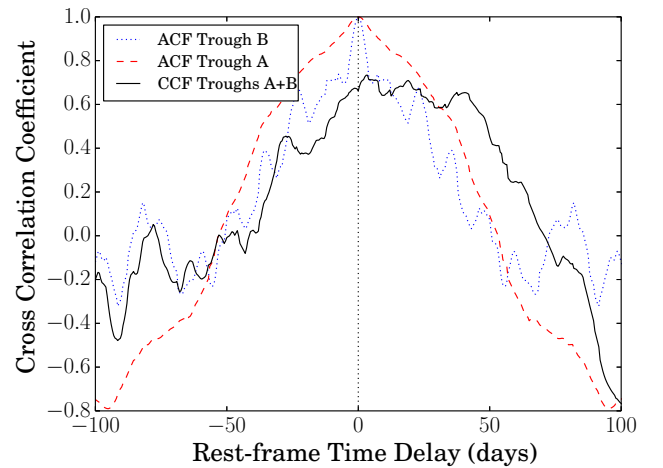}
\caption{The cross correlation function (CCF; black solid line) between the EW light curve of Trough A and the EW light curve of Trough B, showing the Pearson's correlation coefficient as a function of time delay. The auto-correlation function (ACF), which shows the high-velocity trough EW light curve correlated with itself, is show for Trough A (red dashed line) and Trough B (blue dotted line). To eliminate contributions from the \civ \ emission line, these calculations were done using light curves made from the PCA-normalized spectra. }
\label{fig:ccf} 
\end{center}
\end{figure}

As expected based on the RMS of the PCA-normalized spectra, there is no significant variability of Trough C, but the EW light curve for Trough B shows a similar behavior to that of the light curve of Trough A (though with smaller fractional variability). The EW light curves for Trough A and B from the PCA-normalized spectra have a Spearman rank correlation coefficient of 0.525 with a $p$-value of 0.002, indicating that the two are correlated. The cross-correlation function for the two light curves (Figure~\ref{fig:ccf}), showing the Pearson coefficient as a function of time delay between the two light curves,  similarly reveals that they are well-correlated, with a peak $r$ $\sim$ 0.7. We are unable to measure any time delay between the two troughs with sufficient precision to be useful; we measure a time delay of 17~$\pm$~19 days, which is consistent with zero. 

\subsubsection{Short-Term EW Variability} 
\label{sec:shorttermvariability}
\begin{figure*}
\begin{center}
\includegraphics[scale = 0.5, angle = -90, trim = 0 0 0 0, clip]{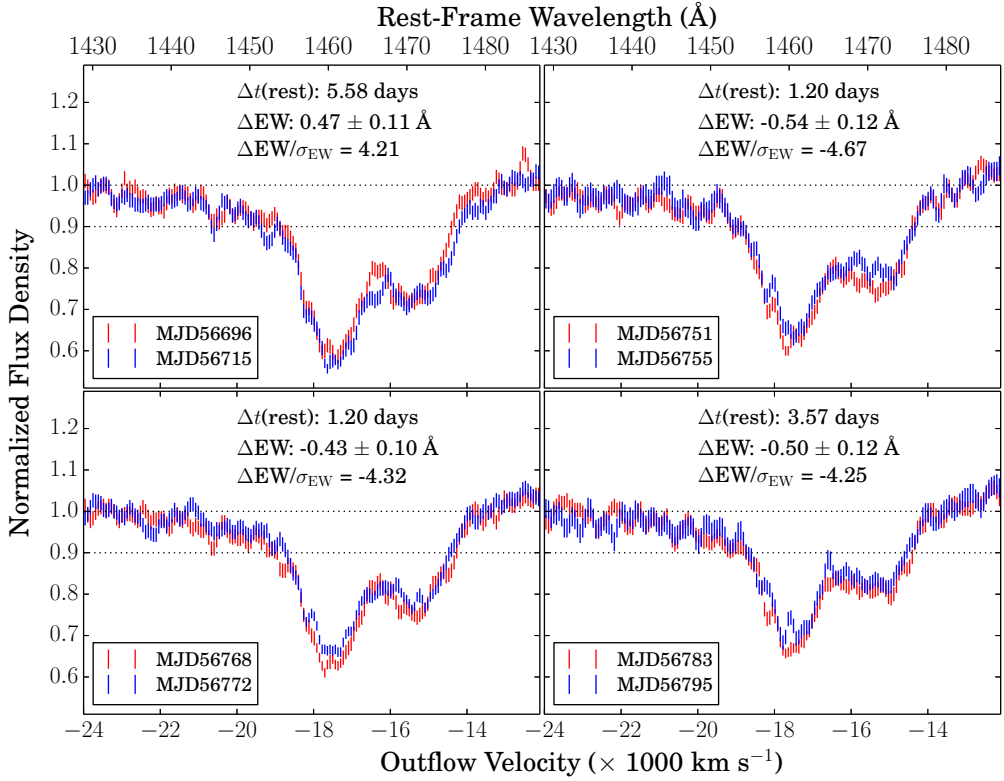}
\caption{Four different pairs of continuum-normalized spectra between which the EW of Trough A varies at greater than 4$\sigma$ significance ($\Delta$ EW $\geq 4 \times \sigma_{\Delta {\rm EW}}$). The continuum-normalized spectra are indicated in red and blue, and the horizontal dashed lines are at a normalized flux density of 1.0 and 0.9 to help guide the eye. The spectra have been smoothed for display purposes (only in this figure) using a boxcar window over 3 pixels to reduce noise. The vertical bars show the 1$\sigma$ uncertainties associated with each point in the smoothed, normalized spectra.}
\label{fig:real_variability}
\end{center}
\end{figure*}

To search for short-timescale variability between epochs in Trough A, we calculated the change in EW ($\Delta$EW) between each pair of sequential observations using the continuum-normalized spectra. Uncertainties in $\Delta$EW ($\sigma_{\Delta {\rm EW}}$) were calculated by summing the uncertainties of the two contributing EW measurements in quadrature. Following \cite{Capellupo13}, in order to be considered ``real" variability, we require a significance of 4$\sigma$ (that $\Delta$EW is at least four times as large as $\sigma_{\Delta {\rm EW}}$). 
We identified four instances of secure variability between sequential epochs during the SDSS-RM campaign, with variability on timescales down to 1.20 days ($\sim$ 29 hours) in the quasar rest frame. Figure~\ref{fig:real_variability} displays the four variable pairs of spectra. 

\begin{figure}
\begin{center}
\includegraphics[scale = 0.3, angle = -90, trim = 0 0 0 0, clip]{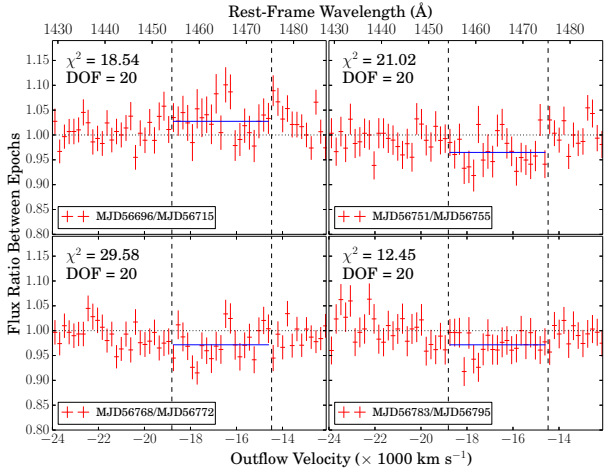}
\caption{Flux ratios between the four different pairs of continuum-normalized spectra shown in Figure \ref{fig:real_variability}. The spectra have been binned by 3 pixels to reduce noise. Red vertical bars show the 1$\sigma$ uncertainty in the flux ratio in each binned pixel, and the horizontal bars show the size of the bins. Blue horizontal lines show the weighted mean of the flux ratio across the Trough~A region.  The $\chi^2$ of the fit and the number of degrees of freedom (DOF) are shown in each panel. In all cases, a fit to this constant model is statistically acceptable.}
\label{fig:fluxratios}
\end{center}
\end{figure}

To assess if this short-term variability in the trough is consistent with coordinated variations across the entire trough, we took the ratio between each pair of spectra and determined whether the flux ratio in the Trough A region is consistent with a constant to within the measurement uncertainties. The flux ratios are shown in Figure \ref{fig:fluxratios} along with a constant fit to the region across the trough. In all cases, a constant model fits the flux ratio in the trough region quite well. However, to be sure that the two sub-troughs within Trough A are not varying differently, we measured individual EW light curves for the blue half and red half of Trough A (which can be decomposed into two BALs of widths at least $\sim2470$ \kms \ and $\sim$1870 \kms, respectively, when ignoring the overlapping wings). We find that the two sub-troughs show the same behavior; i.e., when the blue half is increasing in EW, so is the red half, and by similar fractional changes in EW. Thus, the trough is varying in a very similar, or coordinated, fashion across all velocities.

\begin{figure}
\begin{center}
\includegraphics[scale = 0.47, angle = -90, trim = 0 0 250 300, clip]{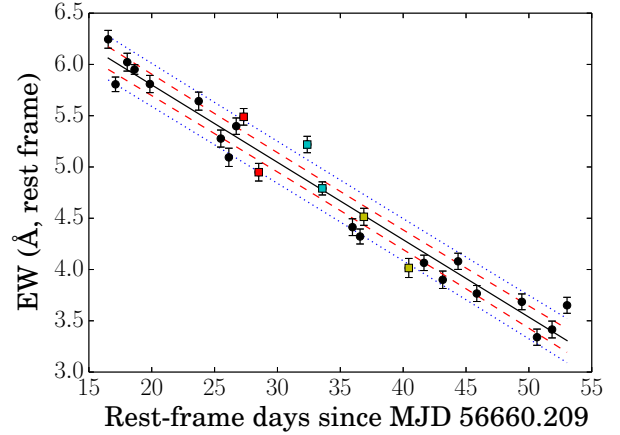}
\caption{The decreasing part of the EW light curve for Trough A (black filled circles) with a linear fit using the \cite{Kelly07b} method that includes intrinsic scatter. The three pairs of epochs between which we see significant variability are shown as colored squares (red, cyan, and yellow). The black solid line shows the linear fit, blue dotted lines show the measured intrinsic scatter, and red dashed lines show the expected scatter about the linear fit assuming no intrinsic scatter; i.e., if the expected scatter was only due to the uncertainties in EW. The expected scatter due to uncertainties in EW is smaller than that observed.} 
\label{fig:lineartrend}
\end{center}
\end{figure}

As discussed in Section \ref{sec:baltroughproperties}, examination of the EW light curve of Trough A (Figure~\ref{fig:lightcurves}) shows a steady increase in EW from rest-frame days 0--15, followed by a steady decreasing trend for the remainder of the campaign. This behavior raises the concern that our observed short-term variability could be merely caused by statistical noise about this long-term trend, which would affect the interpretation of the variability. Three of the four pairs of variable epochs are found in the decreasing half of the light curve, so to determine whether or not these features are likely to be statistical noise, we fit a linear trend to the light curve between days 15--55. A simple linear fit did not perform particularly well (the fit has a reduced $\chi^{2}$ of 5.02), so we instead fit the linear trend allowing for an intrinsic scatter component. We use the Bayesian linear-regression model of \cite{Kelly07b} to fit the line, and find an intrinsic scatter of 0.18 \AA, which is twice the typical measured uncertainty in EW. Explaining this scatter without intrinsic variability would require that our EW uncertainties are underestimated by at least a factor of two: an error this size is unlikely because we account for both pixel uncertainties and noise introduced by the continuum-fitting procedure (see Section \ref{sec:confits}). We present the linear fit, intrinsic scatter, and uncertainties in the fit in Figure~\ref{fig:lineartrend}. 

To explore further the likelihood that our observed short-term variability is due to genuine variability, we created a simulated EW light curve for Trough A, placing each epoch directly on the fitted linear trend and assigning uncertainties equal to our measured uncertainties at each epoch. We then allowed the EW in each epoch to deviate by a random Gaussian deviate reflecting these uncertainties, and counted the number of epoch pairs that deviated from each other by greater than 4$\sigma$ solely due to statistical fluctuations. This simulation was designed to mimic our actual search for pairs of epochs that varied in our data. We repeated this test $10^6$ times to determine the probability that the observed variations are due simply to statistical fluctuations about the long-term decreasing trend. The simulations indicate a probability of only 0.025\% for statistical noise to produce three pairs of sequential EWs that differ by at least 4$\sigma$. When we restrict the simulation to count only those pairs with $\Delta~t<$~2 days in the rest-frame (of which we found two pairs in our light curve), the probability drops to 0.0076\%. Thus, it is highly unlikely that the detected variations are due to statistical fluctuations about the long-term trend, and we conclude that they represent real short-term intrinsic variability in this system.   

\subsubsection{Trough-Profile Variability} 
\label{sec:profilevariability} 
\begin{figure}
\begin{center}
\includegraphics[scale = 0.3, angle = -90, trim = 0 0 0 10, clip]{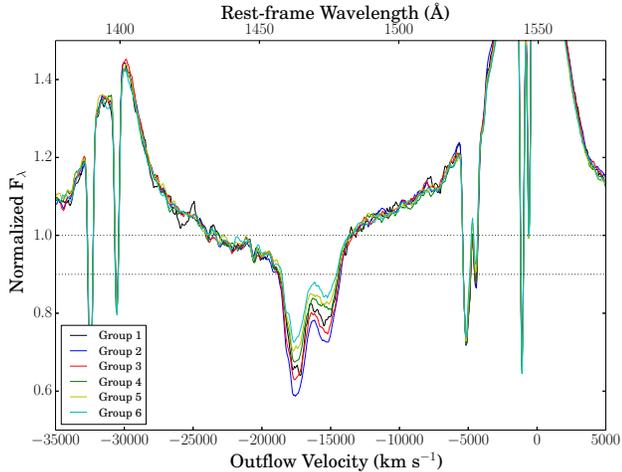}
\caption{Mean continuum-normalized spectra in each group as defined in Section \ref{sec:profilevariability} of the text. Each color shows the mean of each group of spectra throughout the campaign. Horizontal dotted lines are normalized flux densities of 1.0 and 0.9 to guide the eye. The profile of Trough A remains the same throughout the campaign; only the strength/depth of the trough varies.  }
\label{fig:avg_by_ew}
\end{center}
\end{figure}

Though the variations in trough-profile shape are minimal on the shortest timescales, we also search for indications of variability in trough-profile shape associated with the underlying longer-term trends seen in EW in Trough A. To do this, we divided spectra into groups containing spectra that were close together in time and had similar EW measurements. We averaged the continuum-normalized spectra in each group to improve the S/N, allowing us to compare the average trough profiles at different times throughout the campaign. Grouping the spectra resulted in six groups, each containing 3--5 sequential spectra with similar EW values. Group 1 contains the spectra taken from MJD 56660--56669, Group 2 between MJD 56717--56722, Group 3 between MJD 56745--56752, Group 4 between MJD 56780--56783, Group 5 between MJD 56795--56808, and Group 6 between MJD 56813--56837. We averaged all of the normalized spectra in each group and plot each mean spectrum in Figure~\ref{fig:avg_by_ew}. The average trough profiles of each group appear similar throughout the campaign; it is only the strength of the trough that appears to change. To test this conclusion, we measured the flux ratios between each pair of group-averaged spectra at each pixel of the trough. We find that in all cases, a constant model fit to the flux ratio between the different group-averaged spectra is again a statistically acceptable fit, indicating that the absorption trough is changing in a coordinated manner across its entire velocity span on longer timescales as well.

\subsubsection{Emission-Line Variability} 
\label{sec:emissionlinevar}
\begin{figure}
\begin{center}
\includegraphics[scale = 0.43, angle = -90, trim = 0 0 200 200, clip]{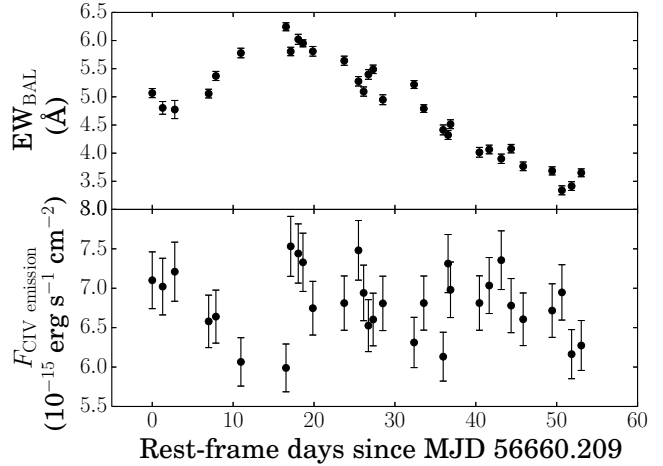}
\caption{The rest-frame EW of the \civ \ BAL (Trough A; top panel) and the integrated flux of the  \civ \ emisson line (lower panel) as a function of rest-frame days since the beginning of the SDSS-RM campaign. The emission-line flux was calculated using only the red half of the \civ \ emission line, which is free of absorption features.} 
\label{fig:civ_emission}
\end{center}
\end{figure}

The SDSS-RM campaign also offers the opportunity to search for links between emission-line and BAL-trough variability and thus possible connections between the high-ionization broad-line region and BAL outflows; such connections have been suggested by many authors (e.g., \citealt{Chiang96}; \citealt{Richards11}; \citealt{Higginbottom13}). As such, we also measured the integrated flux within the \civ \ emission line during the SDSS-RM campaign, shown in Figure~\ref{fig:civ_emission}, to look for possible coordinated variability. We calculated the flux in only the red half of the \civ \ emission line ($\lambda_{\rm rest}~>~1548$~\AA), as the blue half is affected by the mini-BAL troughs B and C. The \civ \ emission-line light curve shows some low-level variations throughout the SDSS-RM campaign, changing by about a factor of $\sim$10\% during this time period, but no large-scale changes similar to those shown by the BAL trough. However, the uncertainties are non-negligible due to the uncertainty of the spectrophotometric flux calibrations (\citealt{Shen15}). 

Based off of its luminosity at rest-frame 1350 \AA \ (as measured from the mean spectrum) and the \civ \ $R-L$ relation of  \cite{Kaspi07}, the expected mean timescale for \civ \ to respond to ionizing-flux changes is on the order of 130 rest-frame days, though gas nearer to the line of sight will respond more quickly.  Because of the uncertainties in integrated emission-line fluxes and because the baseline of the current data from the SDSS-RM campaign is less than $\sim$130 days, we are not able to explore fully the \civ \ emission-line variability --- it might simply not have responded to the observed ionizing continuum yet. Additional observations of this target, with the extended time baseline covered by the ongoing monitoring periods in 2015 and 2016, could be used to explore this further. 

We also consider the \heii \,$\lambda$1640 ($n=3-2$) emission line as a potential probe of the ionizing continuum above 54~eV.
The \heii \ emission line is present but weak in our spectra; we measured the \heii \ EW using the method of \cite{Baskin13}
for ease of comparison with their dataset. Our quasar's average \heii \ EW of $3.5\pm 0.4$ \AA\ is consistent with their BAL quasar average of 3.9\,\AA, but lower than their non-BAL quasar average of 5.1\,\AA\ (their Table 2). From the pre-campaign BOSS spectrum to the last SDSS-RM spectrum, \heii \ shows a marginal weakening in its rest-frame EW of approximately $20\pm 10$\%. The weakness of the \heii \ emission line and the resulting large uncertainties in measuring its EW make it impossible for us to measure short-term variability, and thus the \heii \ emission in this target is not instructive as a probe of the short-term variability of the ionizing continuum. 

\section{DISCUSSION} 
\label{sec:discussion} 
\subsection{Observed Variability in Context} 
\label{sec: obsvar}
Our observations of \civ \ BAL variability in this quasar are the fastest reported detections of significant BAL variability to date. Previously, \cite{Capellupo13} reported variability on timescales down to 8--10 days in two quasars; our shortest observed timescale is shorter by more than a factor of five. 
The targets observed by \cite{Capellupo13} showed variability in only small portions of their respective troughs; we here see coordinated variations across the entire velocity range of the trough, with no discernible BAL-profile changes throughout the campaign. We see very little significant variability in the rest-frame UV continuum as measured by the $i$-band photometric light curve and within the line-free regions of the spectra.  

To determine whether this particular quasar and the \civ \ BAL are atypical when compared to the general BAL population, we compare their various properties with those of the BAL quasar sample of \cite{Gibson09b}. Our target has an absolute magnitude and redshift that are fairly typical of the BAL population; however, the quasar itself is less reddened than typical BAL quasars. The BAL trough itself is somewhat weak, with a smaller EW than many \civ \ BAL troughs, and the higher-velocity portion of the trough is deeper than the low-velocity portion --- this shape is more common in weaker BALs. The velocity width of the trough, at $4340$~ \kms, is not unusual, but is somewhat larger than the majority of \civ \ BAL troughs. The maximum velocity of the trough falls within the normal range of the broader BAL population, but the minimum velocity of the trough falls towards the high end of the distribution (the trough is well-separated from the \civ \ emission-line). In summary, neither the quasar nor its BAL are strong outliers in any of the properties examined, though in some cases they do fall in less common regions of the distribution among BAL quasars. 

\begin{figure}
\begin{center}
\includegraphics[scale = 0.32, angle = -90, trim = 0 0 85 0, clip]{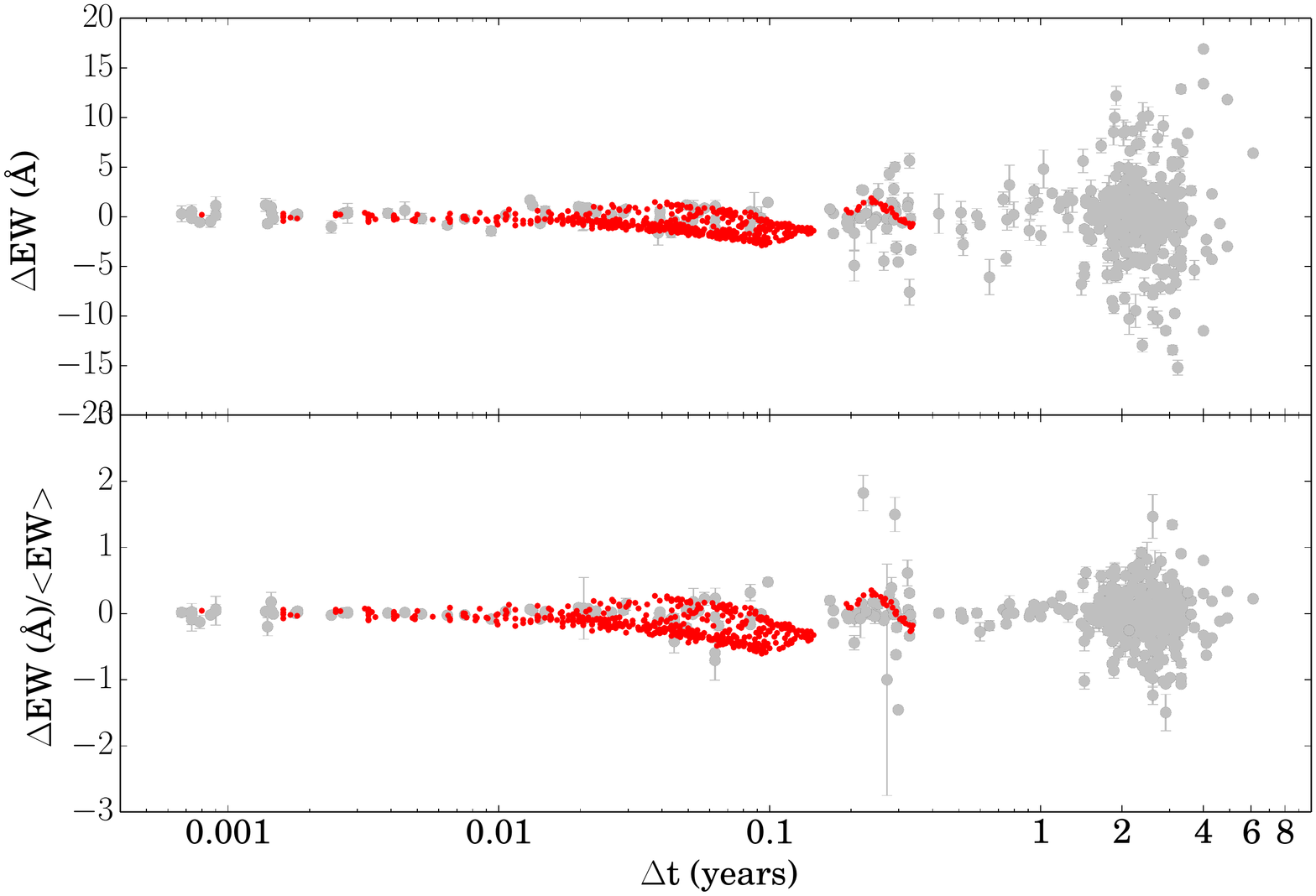}
\caption{Change in EW ($\Delta$EW; top panel) and fractional change in EW ($\Delta$EW/$<$EW$>$; lower panel) as a function of time between epochs ($\Delta t$), reproduced from \cite{Filizak13}. Gray points show data from \cite{Filizak13}, \cite{Barlow93}, \cite{Lundgren07}, and \cite{Gibson08}. Red squares show data for each SDSS-RM epoch paired with every other epoch during the campaign. Values and uncertainties for the SDSS-RM data were calculated following the prescriptions defined in Section 3.3 of \cite{Filizak13}. } 
\label{fig:filizak}
\end{center}
\end{figure}

We also investigate whether the amplitude and timescale of variability seen during the SDSS-RM campaign is normal for a BAL quasar by comparing our observed variability with that of the large sample of \cite{Filizak13}. Figure~\ref{fig:filizak} shows the change and fractional change in EW as a function of $\Delta~t$, reproduced from \cite{Filizak13} with our data overplotted in red. Our points lie within the envelope defined in that work, suggesting that the variability properties seen in our target may be common in quasars; our target is not an apparent outlier in its variability characteristics.  The variability amplitudes on short timescales during our campaign are also consistent with those observed on short timescales in a few targets by \cite{Capellupo13}. 

There are a number of potential causes of BAL variability previously discussed in the literature, and the physical constraints we can place on the nature of the outflow are dependent on the mechanism driving the variability. Below we briefly consider a few possible causes of the observed BAL variability and use our measurements to determine the relevant physical parameters of the outflow. 

\subsection{Possible Causes of Variability} 
\label{sec:causes} 

\subsubsection{Ionization State Variability}
\label{sec:ionization_variability}
Recent studies by Filiz Ak et al. (2012, 2013) have demonstrated that, in general, distinct \civ \ BAL troughs in the same object separated by as much as 10,000--20,000 \kms \ tend to strengthen or weaken together. Such correlated changes over large velocity ranges suggest that much BAL variability arises from changes in the ionizing flux reaching the absorber rather than from transverse gas motions. There are several aspects of our observations that suggest variations in the ionization state of the gas are responsible for the observed variability. First, global changes occur across the entire trough rather than in small segments --- this can be understood if the ionization state of the gas changes but is difficult to explain in other scenarios (see below). Secondly, we see coordinated variability between the high-velocity \civ \ BAL Trough A and the mini-BAL Trough B (Figures~\ref{fig:pca_ewlightcurves} and \ref{fig:ccf}), which are separated by $\sim$12,000 \kms. 

A notable aspect of our observations is the lack of significant variability in the rest-frame UV continuum flux, as measured in the photometric light curves (Figure~\ref{fig:phot_lcs}); at most, the photometric flux in our observations varies by only 10\%. However, the flux in the $i$ band is not necessarily a reliable tracer of the ionizing continuum --- \civ \ is ionized by radiation in the extreme ultraviolet (EUV) part of the spectrum, which is likely more variable than the observed emission. For example, \cite{Marshall97} monitored the Seyfert 1 galaxy NGC\,5548 using the {\it Extreme Ultraviolet Explorer}, finding that the EUV continuum at 0.15--0.2 keV was significantly more variable than that of the continuum at 1350 \AA;  at times the EUV was observed to vary by factors of $\sim$2 while the lower-energy UV as observed with the {\it Hubble Space Telescope} varied by only $\sim$10\% during the same period. The EUV was found to vary in phase with the UV/optical continuum to within the time sensitivity limit imposed by the cadence of the observations. 

While there are few other examples of AGN variability monitoring in the EUV (e.g., \citealt{Marshall96}), the same trend has been seen in X-rays;  simultaneous observations in the X-ray and optical have shown that the short-term variations in X-ray flux are often not detected in the optical, though correlations are seen on longer timescales (e.g., \citealt{Peterson00b}). These studies were done with local Seyfert galaxies rather than distant quasars; however, the X-ray variability properties are broadly similar among the two populations --- both quasars and nearby Seyferts show strong, rapid X-ray variability (e.g., \citealt{Gibson12}; \citealt{Shemmer14}). Such strong variations of the ionizing continuum could be sufficient to drive the observed rapid ionization response of the BAL trough. Furthermore, if the ionization conditions of the absorber are such that the \civ \ abundance has a steep dependence on the ionization parameter (e.g., see Figure 1 of \citealt{Kallman82}), even more modest changes in the ionizing continuum can drive strong changes in the \civ \ column density along the line of sight. 

Clumpiness of the X-ray ``shielding gas" (from the model of \citealt{Murray95}) is another possible explanation for the observed lack of photometric variability. In this model, the shielding gas is located between the continuum source and the outflow, and prohibits high-energy radiation (soft X-rays and potentially some of the ionizing continuum) from reaching the outflow, but does not necessarily affect the lower-energy observed UV continuum. If the shielding gas is corotating with the disk, clumpiness or movement of the shielding gas could cause the amount of ionizing flux reaching the outflow to vary on shorter timescales than the ionizing continuum changes themselves. Models do predict variability on short timescales related to the shielding gas (e.g., \citealt{Proga00}; \citealt{Sim10}) and a few studies report possible detections of this as well (e.g., \citealt{Gallagher04}; \citealt{Saez12}). If our BAL variability is driven predominantly by changes in the shielding gas, then the rapidity of the trough variability constrains models of the size and kinematics of the shielding gas rather than the outflow gas. 

\subsubsection{Cloud Crossing}
\label{sec:cloudcrossing}
Another potential source of variability is the transverse movement of the outflow material across our line of sight to the quasar source (``cloud crossing"). Several previous studies of BAL variability are consistent with this scenario, as variability is often observed in only narrow portions of the BAL troughs (e.g.,  \citealt{Hall11}; \citealt{Vivek12}; \citealt{Capellupo13}). However, for the \civ \ BAL trough in our target, which is $\sim$4300~km\,s$^{-1}$ wide, to undergo coherent strengthening or weakening across its entire velocity span  as observed would require the absorbing gas at all velocities to be physically connected. This explanation is implausible, as the trough width is too large to be sustained as a turbulent velocity dispersion without dispersing the absorbing structure 
(e.g., see \S 5.2 of \citealt{Rogerson11})
or heating it to a temperature at which no \civ \ would be seen (the thermal velocity in photoionized gas exhibiting \civ \ absorption will be $\leq 10$~km\,s$^{-1}$). An absorbing structure with a gradient in line-of-sight velocity across the continuum source can explain the trough width, but not the coherent variability as a function of velocity, because gas at different velocities will cover different regions of the continuum source and will not be moving into or out of the same line of sight at the same time, as required to explain our observations.  We can think of no plausible absorbing structure whose bulk motion, either into or out of our line of sight to the entire continuum emission region (or to hot spots within that region), can reproduce our observations. Furthermore, this scenario does not explain the coordinated variability seen between Trough A and Trough B. 

\subsubsection{Inhomogeneous Emission}
\label{sec:inhomogeneous}
Partial coverage of a variable inhomogeneous accretion disk (e.g., \citealt{Dexter11}) by an absorber can cause EW variability in the resulting trough. If the part of the disk not covered by an absorber brightens, both the observed continuum and the residual flux in the trough will increase, and the normalized trough will appear to weaken; conversely, if that part of the disk fades, the EW of the trough will increase due to the fading continuum. If the part of the disk covered by an absorber brightens, the observed continuum flux will increase more than the residual flux in the trough will, and the normalized trough will appear to strengthen; conversely, if that part of the disk fades, the normalized trough will appear to weaken.
However, in any of the above cases, for the mean fractional depth of the trough to decrease smoothly by 12\% of the continuum level from rest-frame day 18 to 53, as observed, would require the observed continuum flux to change smoothly by a similar percentage in that time frame. The photometric light curves suggest no such trend. Therefore, we do not consider an inhomogeneous emitter a viable explanation for the observed variability.

\subsection{Physical Constraints from the Lack of Trough-Profile Variability} 
\label{sec:noprofilevar} 
We found in Section \ref{sec:profilevariability} that Trough A is consistent with having an unchanged line profile throughout the campaign; only its strength varies. The normalized flux density at velocity $v$ in a trough can be characterized as $F_v~=~(1-C_v)~+~C_ve^{-\tau_v}$, where $C_v$ is the fraction of the emission source covered by the absorber and $\tau_v$ is the optical depth in the absorbing ion, both along our line of sight.
In cloud crossing variability, changes in $C_v$ predominate; in ionization variability, changes in $\tau_v$ predominate.
The S/N of our spectra are not high enough to distinguish statistically between changing $C_v$ while keeping $\tau_v$ fixed at all $v$ and changing $\tau_v$ while keeping $C_v$ fixed at all $v$; both models are acceptable fits to our data.
However, the former scenario requires coordinated changes in $C_v$ across the entire width of the profile, for which we found no plausible mechanism in Section \ref{sec:cloudcrossing}.
On the other hand, the latter scenario is naturally explained by ionization variability.
Adopting that scenario, we can constrain the gas to be at most marginally optically thick, with $\tau_{\rm max}=0.55$ (for constant $C_v=1$) to $\tau_{\rm max}=1.5$ (for constant $C_v=0.47$), and with gas at most velocities and in most epochs having $\tau_v$ well below $\tau_{\rm max}$.

\subsection{Physical Constraints from Ionization State Variability}
\label{sec:ionization_constraints}

Following the discussion in Section \ref{sec:causes}, we conclude that a change in ionization state of the absorbing gas is the most likely cause of the observed BAL variability in our target. Regardless of whether the change in ionizing flux reaching the absorber is caused by fluctuations in the high-energy ionizing continuum itself, or due to changes in shielding gas, we can use our observations to place constraints on the outflow properties. 
For gas which is initially in photoionization equilibrium,
the characteristic timescale for changes in the density $n_i$ of ionization stage $i$ of some element in response to an ionizing flux change can be written as
\begin{equation}
t_i^* = \left[ -f_i \left(I_i-n_e\alpha_{i-1}\right) \right]^{-1}
\end{equation}
(equivalent to Eq.\ 10 of \citealt{Arav12}),
where $-1<f_i<+\infty$ is the fractional change in
$I_i$, the ionization rate per ion of stage $i$ ($I_i(t>0) = (1+f_i)I_i(t=0)$), $\alpha_{i-1}$ is the recombination coefficient to ionization stage $i-1$ of the ion,
and a negative timescale represents a decrease in $n_i$ with time.
This equation only considers photoionization and recombination; other processes are neglected.
Assuming an absorber of constant thickness, observed changes in ionic column density
(i.e., $\tau$) correspond to ionic density changes.  Gas showing such variations is not in a steady state by definition, but such gas can still be in equilibrium with a varying ionizing flux if its $t_i^*$ is considerably shorter than the flux variability timescale (e.g., \S 6 of \citealt{Krolik95}).
For optically thin gas at distance $r$ from a quasar with luminosity $L_\nu$ at frequency $\nu$, the ionization rate per ion of stage $i$ is given by
\begin{equation}
\label{eq2}
I_i=\int_{\nu_i}^{\infty}\frac{(L_\nu/h\nu)\sigma_\nu}{4\pi r^2} d\nu
\end{equation}
where $\sigma_\nu$ is the ionization cross-section for photons of energy $h\nu$. 

If the absorbing gas is sufficiently far from the quasar that
$I_i\ll n_e\alpha_{i-1}$, then Equation 1 reduces to $t_{\rm rec}~=~1/f_in_e\alpha_{i-1}$ (which is just the recombination time of the ion in the $f_i=-1$ case where the ionizing flux drops to zero; e.g., \citealt{Capellupo13}),
and the observed absorption variability timescale constrains the density of the absorber.  
However, if the absorbing gas is close enough to the quasar that
$I_i\gg n_e\alpha_{i-1}$, then the relevant timescale is $t_i=-1/f_iI_i$
and the absorption variations of the ion reflect the ionizing flux variations of the quasar, with no density constraint derivable just from absorption variations.
An observed timescale for variations in optically thin absorption therefore constrains the absorbing gas to either have a density $n_e>n_{\rm min}$ and $r>r_{\rm equal}$, where $r_{\rm equal}$ is the distance at which $I_i=n_{\rm min}\alpha_{i-1}$, or to be located at $r<r_{\rm equal}$ with no constraint on the density.

To determine the constraints in our target, we assume a temperature of $\log {T} = 4.3$, the temperature of the broad line region gas in the \civ \ region (e.g., \citealt{Krolik99}).  
The recombination coefficient corresponding to this temperature is 
$\alpha_{\rm C III}~=~2.45~\times~10^{-11}$~cm$^3$~s$^{-1}$ (from the CHIANTI online database; \citealt{Dere97}; \citealt{Landi13}). 
For the simple case of the ionizing flux dropping to zero, $f_i = -1$ and the timescale $t_i^*$ above can be approximated as the recombination time $t_{\rm rec}$ $\sim$ 1/$n_{e}\alpha_{C{\sc III}}$. Using our observed timescale of 1.20 days as an upper limit to $t_{\rm rec}$ for Trough A, we calculate a lower limit on the density of the gas to be \density.
We cannot search Trough B for short-term variations as we can in Trough A, because Trough B sits on top of a slowly varying, imperfectly modeled emission line. However, we can set a lower limit on $n_e$ for Trough B using the observed EW decrease of about 20\% in 20 days. This indicates a characteristic timescale of $\sim$100 days and a minimum $n_e$ about 100 times lower than that in the Trough A absorber.

Our calculation for Trough A at log~$T$~=~4.3 yields similar densities to those found by \cite{Capellupo13} for their targets, which showed variability on timescales down to 8--10 days, though they used a different recombination coefficient (presumably calculated for a different temperature). Using the same recombination coefficient they used with our observed timescale would yield $n_e$ for our target higher than their measurement by more than a factor of six ($n_e$~\gtsim~2.7~$\times~10^6$~cm$^{-3}$). 

Using our lower limit density of \density, we calculate the minimum distance from the quasar at which that lower limit is valid.
From its observed flux density at rest-frame 3000\,\AA, our quasar has a bolometric luminosity $L_{\rm Bol}~=~1.7\times~10^{47}$~erg~s$^{-1}$ using a bolometric correction factor of 5 (\citealt{Richards06}). We adopt the UV-soft spectral energy distribution of \cite{Dunn10}  to calculate $L_\nu$ and integrate Equation \ref{eq2}, setting $I_i$ = $n_e$$\alpha_{\rm C III}$ to solve for the radius $r_{\rm equal}$ at which the ionization rate is equal to the recombination rate. We find $r_{\rm equal}=3.8\times 10^{20}$ cm (120 pc).
Thus, the absorber in this quasar either has \density \ and $r>~120$~pc, or it has $r~<~120$~pc, in which case the recombination timescale cannot be used to constrain the density. This radius is much larger than the launching radius proposed by theoretical studies, which is on the order of 10$^{-3}$ pc (e.g., \citealt{Murray95}; \citealt{Proga00}; \citealt{Higginbottom13}) --- some also theorize that BALs are launched at distances comparable to the radius of the broad line-emitting region (e.g., \citealt{Baskin14}). A physical mechanism for launching such high-velocity outflows at distances of parsecs or kiloparsecs that is consistent with observations is difficult to produce. However, $r ~>$~120~pc falls within the wide spread of outflow radii reported in the literature --- many studies report outflow radii on the order of parsecs or kiloparsecs (see, e.g.,  Table 10 of \citealt{Dunn10} and references therein), though some studies do report significantly smaller radii (e.g., \citealt{Leighly09}; \citealt{Capellupo13}). Our measured lower limit on $n_e$ is higher than those typically found for the outflows at kpc-scale distances, which are on the order of $10^3-10^4$~cm$^{-3}$ (e.g., \citealt{Dunn10}; \citealt{Hall11}; \citealt{Borguet13}), though we make this comparison with caution due to the differences in analysis methods and possible biases within the samples of outflows. Our density lower limit is closer to other measurements of outflows that are found at distances on the order of $r_{\rm equal}$ ($\sim$100 pc), which have $n_e$ $\sim10^4-10^5$~cm$^{-3}$ (e.g., \citealt{Arav13};  \citealt{Borguet13}). 

The discrepancy between observations and model predictions can likely be reconciled through selection effects that lead to at least some BAL outflows (particularly those with distance measurements to date) being observed at radii much larger than the launching radii (e.g., \citealt{Lucy14}). The radius at which BAL outflows are observed is not necessarily the radius from which they are launched; some studies suggest that outflows can maintain high velocities at large distances allowing us to observe them much farther out than the launching location (e.g., \citealt{Faucher12}). 
In fact, we can use our observations to estimate the upper limit for the launching radius $r_{\rm launch}$ of the outflow; however, to do so requires the adoption of an outflow model. For example, within the context of the radiation pressure confinement (RPC) model of \cite{Baskin14}, the terminal velocity of an outflow is $v_{\rm term} \sim v_{\rm Kep} \sqrt{\Gamma L/L_{\rm Edd}}$, where $\Gamma$ is the force multiplier and $v_{\rm Kep}$ is the circular velocity of gas at the launch radius (e.g., \citealt{Laor02}). Adopting  $\Gamma \sim$ 10 from \cite{Baskin14} and assuming that our BAL is observed at terminal velocity, we find $r_{\rm launch} < 0.3$ pc $\sim$ 400 light-days. The mean radius of the \civ-emitting broad line region in this quasar is estimated to be $R_{\rm BLR}$ $\sim$ 130 light-days (see Section \ref{sec:emissionlinevar}), and with $M_{\rm BH} = 1.1 \times 10^{9}$~\Msun, this target has a Schwarzchild radius $R_{\rm Sch} \sim$ 0.125 light-days. Thus for this quasar, assuming the RPC model of \cite{Baskin14} yields $r_{\rm launch} <$ 3$R_{\rm BLR}$ $\sim$ 3200$R_{\rm Sch}$. Note that $r_{\rm launch}$ is model dependent and thus different values will be obtained depending upon the model.

We can also place constraints on $n_e$ using the presence of other absorption features, or the lack thereof. For example: With a density of \density \ at $r~=~r_{\rm equal}$, the gas would have an ionization parameter of $U_H~=~0.07$ for our assumed SED ($U_H~=~Q_H/4\pi r^2 n_Hc$, with $Q_H~=~1.2\times~10^{57}$ hydrogen-ionizing photons s$^{-1}$ and $n_H~=~0.82n_e$).  That value of $U_H$ is consistent with the presence of absorption from \civ \ but not C\,{\sc ii}\,$\lambda$1335 in this object. A factor of $\sim$50 lower $U_H$ would yield C\,{\sc ii}\,$\lambda$1335 absorption about half as deep as that of \civ \ (Fig.\ 3 of \citealt{Hamann95}), 
which is not observed in this source. Given that $U_H$ can be lower by at most a factor of 50, if the gas is located at $r>r_{\rm equal}$, it is thus also constrained to have  $r_{\rm equal} \leq r \lesssim \sqrt{50}r_{\rm equal}$ and also must have density $3.9\times 10^5 \leq n_e \leq 10^7(r_{\rm equal}/r)^2$ cm$^{-3}$. 

We next measure the kinetic luminosity of the outflow relative to its Eddington luminosity to evaluate the potential of the outflow to provide substantial feedback to the quasar host galaxy. To estimate the Eddington luminosity ($L_{\rm Edd}$) of this quasar, we first require an estimate of $M_{\rm BH}$. We use Equation 9 from \cite{Rafiee11}, which incorporates the dispersion $\sigma$ (in \kms) of the Mg\,{\sc ii} line: 
\begin{equation} 
\label{eq:mbh}
M_{\rm BH}/M_\odot = 30.5 [\lambda L_{3000}/10^{44}] ^{0.5} \sigma^2 
\end{equation} 
 We assume $\sigma~=~$FWHM/2.35, with FWHM~=~3274~\kms \ (\citealt{Paris15}). 
The observed flux density at rest-frame wavelength 3000\,\AA \ is $f_{3000}$~=~8$\times$10$^{-17}$~\ergscm ~\AA$^{-1}$, which corresponds to 
$\lambda L_{3000}$ = $3.38\times 10^{46}$ erg s$^{-1}$. 
Using the above equation yields $M_{\rm BH} = 1.1 \times 10^{9}$~\Msun \
with an intrinsic scatter of $0.4\times 10^{9}$ (\citealt{Rafiee11}). 
With our calculation of \mbh above, we measure $L_{\rm Edd}~=~1.7\times 10^{47}$ \ergs,
and thus in this case the Eddington ratio, $f_{\rm Edd} = L_{\rm bol}/L_{\rm Edd}$, is about 1. We note that \mbh measurements made from single spectra such as this are subject to a number of uncertainties; we consider our measured $L_{\rm Edd}$ to be an approximation. 

The kinetic luminosity of a BAL outflow consisting of a thin shell at distance $r$ from the quasar is
\begin{equation} 
L_k=2\pi\mu m_p\Omega r N_H v^3
\end{equation} 
(e.g., Eq. 1 of \citealt{Arav13}) where $\Omega$ is the global BAL covering fraction (unity for a spherical outflow), $\mu m_p=1.4m_p$ is the mean mass per proton in the absorber, and $N_H$ is the absorber's hydrogen column. We assume $C_v=1$ (see Section \ref{sec:noprofilevar}) and $\Omega=0.2$ (derived from the overall fraction of quasars hosting \civ \ BALs; e.g., \citealt{Gibson09b} or \citealt{Dunn10}), and the average \civ \ $\tau_v$ of Trough A is 0.26. The column density per \kms \ for an unsaturated and resolved doublet absorption feature is $N_v~=~(3.7679~\times~10^{14}$~cm$^{-2})\tau_v/\lambda f_{ij}$ (\citealt{Arav99}; \citealt{Hall03}); we assume an unresolved, optically thin trough and use an average $\lambda=1549.055$~\AA\ and a summed oscillator strength $f_{ij}=0.286$. For Trough A, with a total width of 4340 \kms, we have a total $N_{\rm CIV}~=~9.6 \times~10^{14}$~cm$^{-2}$. Conservatively assuming that all carbon is in the form of \civ, then for solar abundance (\citealt{Asplund09}), we have $N_H=3.7\times 10^{18}$ cm$^{-2}$ in the Trough A outflow. With a mean absorbed-flux-weighted centroid velocity of $v=1.675\times 10^9$ cm s$^{-1}$ over the campaign (see Section \ref{sec:baltroughproperties}) and assuming a distance of 120 pc, the Trough A outflow has $L_k/L_{\rm Edd}~=~1.1\times~10^{-4}$, or 0.01\%. 
AGN/host-galaxy feedback models require $L_{k}/L_{\rm Edd}$ \gtsim 0.5\% to 5\% in order for the outflowing material to provide substantial feedback to the host galaxy (e.g., \citealt{Dimatteo05}; \citealt{Hopkins05a}; \citealt{Hopkins10}), which is about an order of magnitude larger than our measurement. Thus, we find no evidence that the outflow will contribute significant feedback to the host galaxy. However, it could do so if it is accompanied by a sufficient column density of very high ionization gas, to which we are not sensitive. 

 \section{SUMMARY AND FUTURE WORK} 
\label{sec:conclusions}
We have discovered and studied a rapidly variable \civ \ BAL trough in a quasar observed as a part of the SDSS-RM campaign using 31 epochs of BOSS spectroscopic observations and accompanying optical photometric light curves obtained over six months in 2014. Our main findings are the following: 
\begin{enumerate} 
\item We measure significant variability of the EW of the high-velocity \civ \ BAL trough (Trough A) on timescales as short as 1.20 days in the quasar rest frame, the shortest timescale yet reported. The EW varied on the order of $\sim$10\% on the smallest timescales, while over the entire campaign, the EW varied by more than factor of two. See Section \ref{sec:balmeasurements}. 
\item The rest-frame UV photometric flux of the quasar shows little variability ($\sim$10\%) during the length of the campaign. See Section \ref{sec:photobs}. 
\item The trough itself shows coordinated variability across its entire velocity range throughout the campaign, with no significant change in shape across its 4340 \kms \ span, suggesting that the variability is not due to bulk motion of the gas. See Sections \ref{sec:profilevariability} and \ref{sec:cloudcrossing}. 
\item We measure coordinated variability between a lower-velocity \civ \ mini-BAL (Trough B) and the highest-velocity \civ \ BAL (Trough A); this behavior, combined with the coordinated variability across the entire Trough A structure, suggests that the variability is driven by a change in ionization state of the gas.  See Sections~\ref{sec:baltroughproperties} and \ref{sec:ionization_variability}. 
\item For a change in ionization state, a variability timescale of 1.20 days suggests a lower limit to the density of the gas of \density \ so long as the gas is at a radius of greater than $\sim$ 120 pc. We also measure the kinetic luminosity $L_k~=~2.2~\times~10 ^{-4}~L_{\rm Edd}$, which suggests that the outflow itself is not a major contributor to host galaxy feedback; however, since this is an approximation only, we do not rule out the possibility. See Section \ref{sec:ionization_constraints}.

\end{enumerate} 

The key to this successful investigation of such short-timescale variability was the high cadence of the SDSS-RM spectroscopic data; this allowed a robust exploration of changes on shorter timescales than have previously been investigated. There are several other BAL quasars that were observed as a part of this program that have not yet been thoroughly investigated; a search for similar short-term BAL variability in these targets will be instructive and will help determine the frequency of this behavior. Additional monitoring of these BAL quasars will be obtained in the following years as a part of the ongoing SDSS-RM program. While the upcoming observations will not provide spectra at the same high cadence that was obtained in 2014, the additional data will allow a search for continued BAL variability in these targets and extend the baseline for the longer-term trends observed (for example, the troughs in this object are present at approximately the same strength in the first few spectra of 2015). In addition, the extended baseline of these observations will allow us to explore the relationship between BAL trough and emission-line variability (see Section~\ref{sec:emissionlinevar}),  which has not previously been investigated due to the lack of suitable data for such a study. 

Figure~\ref{fig:filizak} shows that the variability properties of this target are consistent with those found in other quasars, indicating that short-term variability such as we see may be common. High-cadence, high-S/N spectroscopic observing programs geared toward investigating variability in additional BAL quasars would expand the sample, allow us to constrain further the dominant driving mechanisms of BAL variability in quasars, and refine models describing the environment and physics regulating BALs as well as their possible contributions to feedback to their host galaxies. 

 \acknowledgments 
We thank A. Baskin for valuable discussions with regards to this work and also our anonymous referee for their helpful suggestions. We also thank Vahram Chavushyan for providing us with the original spectrum of this target taken in 1991. CJG and WNB acknowledge support from NSF grant AST-1108604 and the V.M. Willaman Endowment. PBH is supported by NSERC. JRT and YS acknowledge support from NASA through Hubble Fellowship grants \#51330 \#51314, respectively, awarded by the Space Telescope Science Institute, which is operated by the Association of Universities for Research in Astronomy, Inc., for NASA under contract NAS 5-26555. KDD is supported by an NSF AAPF fellowship awarded under NSF grant AST-1302093. CSK acknowledges the support of NSF grant AST-1009756. BMP acknowledges support from NSF Grant AST-1008882 to The Ohio State University. Funding for SDSS-III has been provided by the Alfred P. Sloan Foundation, the Participating Institutions, the National Science Foundation, and the U.S. Department of Energy Office of Science. The SDSS-III web site is http://www.sdss3.org/.

SDSS-III is managed by the Astrophysical Research Consortium for the Participating Institutions of the SDSS-III Collaboration including the University of Arizona, the Brazilian Participation Group, Brookhaven National Laboratory, Carnegie Mellon University, University of Florida, the French Participation Group, the German Participation Group, Harvard University, the Instituto de Astrofisica de Canarias, the Michigan State/Notre Dame/JINA Participation Group, Johns Hopkins University, Lawrence Berkeley National Laboratory, Max Planck Institute for Astrophysics, Max Planck Institute for Extraterrestrial Physics, New Mexico State University, New York University, Ohio State University, Pennsylvania State University, University of Portsmouth, Princeton University, the Spanish Participation Group, University of Tokyo, University of Utah, Vanderbilt University, University of Virginia, University of Washington, and Yale University. 

This work has made use of the CHIANTI Atomic Database for Spectroscopic Diagnostics of Astrophysical Plasmas. CHIANTI is a collaborative project involving George Mason University, the University of Michigan (USA), and the University of Cambridge (UK). 


\begin{thebibliography}{73}
\expandafter\ifx\csname natexlab\endcsname\relax\def\natexlab#1{#1}\fi

\bibitem[{{Alam} {et~al.}(2015){Alam}, {Albareti}, {Allende Prieto}, {Anders},
  {Anderson}, {Andrews}, {Armengaud}, {Aubourg}, {Bailey}, {Bautista}, \&
  et~al.}]{Alam15}
{Alam}, S., {et~al.} 2015, ArXiv e-prints

\bibitem[{{Allen} {et~al.}(2011){Allen}, {Hewett}, {Maddox}, {Richards}, \&
  {Belokurov}}]{Allen11}
{Allen}, J.~T., {Hewett}, P.~C., {Maddox}, N., {Richards}, G.~T., \&
  {Belokurov}, V. 2011, \mnras, 410, 860

\bibitem[{{Arav} {et~al.}(2013){Arav}, {Borguet}, {Chamberlain}, {Edmonds}, \&
  {Danforth}}]{Arav13}
{Arav}, N., {Borguet}, B., {Chamberlain}, C., {Edmonds}, D., \& {Danforth}, C.
  2013, \mnras, 436, 3286

\bibitem[{{Arav} {et~al.}(2012){Arav}, {Edmonds}, {Borguet}, {Kriss},
  {Kaastra}, {Behar}, {Bianchi}, {Cappi}, {Costantini}, {Detmers}, {Ebrero},
  {Mehdipour}, {Paltani}, {Petrucci}, {Pinto}, {Ponti}, {Steenbrugge}, \& {de
  Vries}}]{Arav12}
{Arav}, N., {et~al.} 2012, \aap, 544, A33

\bibitem[{{Arav} {et~al.}(1999){Arav}, {Korista}, {De Kool}, {Junkkarinen}, \&
  {Begelman}}]{Arav99}
{Arav}, N., {Korista}, K. T., {De Kool}, M., {Junkkarinen}, V.T., \& {Begelman}, M. C.
  1999, \apj, 516, 27

\bibitem[{{Asplund} {et~al.}(2009){Asplund}, {Grevesse}, {Sauval}, \&
  {Scott}}]{Asplund09}
{Asplund}, M., {Grevesse}, N., {Sauval}, A.~J., \& {Scott}, P. 2009, \araa, 47,
  481

\bibitem[{{Barlow}(1993)}]{Barlow93}
{Barlow}, T.~A. 1993, PhD thesis, California University

\bibitem[{{Baskin} {et~al.}(2013){Baskin}, {Laor}, \& {Hamann}}]{Baskin13}
{Baskin}, A., {Laor}, A., \& {Hamann}, F. 2013, \mnras, 432, 1525

\bibitem[{{Baskin} {et~al.}(2014){Baskin}, {Laor}, \& {Stern}}]{Baskin14}
{Baskin}, A., {Laor}, A., \& {Stern}, J. 2014, \mnras, 445, 3025

\bibitem[{{Bertin} \& {Arnouts}(1996)}]{Bertin96}
{Bertin}, E., \& {Arnouts}, S. 1996, \aaps, 117, 393

\bibitem[{{Borguet} {et~al.}(2013){Borguet}, {Arav}, {Edmonds}, {Chamberlain},
  \& {Benn}}]{Borguet13}
{Borguet}, B.~C.~J., {Arav}, N., {Edmonds}, D., {Chamberlain}, C., \& {Benn},
  C. 2013, \apj, 762, 49

\bibitem[{{Capellupo} {et~al.}(2013){Capellupo}, {Hamann}, {Shields},
  {Halpern}, \& {Barlow}}]{Capellupo13}
{Capellupo}, D.~M., {Hamann}, F., {Shields}, J.~C., {Halpern}, J.~P., \&
  {Barlow}, T.~A. 2013, \mnras, 429, 1872

\bibitem[{{Cardelli} {et~al.}(1989){Cardelli}, {Clayton}, \&
  {Mathis}}]{Cardelli89}
{Cardelli}, J.~A., {Clayton}, G.~C., \& {Mathis}, J.~S. 1989, \apj, 345, 245

\bibitem[{{Chavushyan} {et~al.}(1995){Chavushyan}, {Stepanyan}, {Balayan}, \&
  {Vlasyuk}}]{Chavushyan95}
{Chavushyan}, V.~O., {Stepanyan}, D.~A., {Balayan}, S.~K., \& {Vlasyuk}, V.~V.
  1995, Astronomy Letters, 21, 804

\bibitem[{{Chiang} \& {Murray}(1996)}]{Chiang96}
{Chiang}, J., \& {Murray}, N. 1996, \apj, 466, 704

\bibitem[{{Dawson} {et~al.}(2013){Dawson}, {Schlegel}, {Ahn}, {Anderson},
  {Aubourg}, {Bailey}, {Barkhouser}, {Bautista}, {Beifiori}, {Berlind},
  {Bhardwaj}, {Bizyaev}, {Blake}, {Blanton}, {Blomqvist}, {Bolton}, {Borde},
  {Bovy}, {Brandt}, {Brewington}, {Brinkmann}, {Brown}, {Brownstein}, {Bundy},
  {Busca}, {Carithers}, {Carnero}, {Carr}, {Chen}, {Comparat}, {Connolly},
  {Cope}, {Croft}, {Cuesta}, {da Costa}, {Davenport}, {Delubac}, {de Putter},
  {Dhital}, {Ealet}, {Ebelke}, {Eisenstein}, {Escoffier}, {Fan}, {Filiz Ak},
  {Finley}, {Font-Ribera}, {G{\'e}nova-Santos}, {Gunn}, {Guo}, {Haggard},
  {Hall}, {Hamilton}, {Harris}, {Harris}, {Ho}, {Hogg}, {Holder}, {Honscheid},
  {Huehnerhoff}, {Jordan}, {Jordan}, {Kauffmann}, {Kazin}, {Kirkby}, {Klaene},
  {Kneib}, {Le Goff}, {Lee}, {Long}, {Loomis}, {Lundgren}, {Lupton}, {Maia},
  {Makler}, {Malanushenko}, {Malanushenko}, {Mandelbaum}, {Manera}, {Maraston},
  {Margala}, {Masters}, {McBride}, {McDonald}, {McGreer}, {McMahon}, {Mena},
  {Miralda-Escud{\'e}}, {Montero-Dorta}, {Montesano}, {Muna}, {Myers},
  {Naugle}, {Nichol}, {Noterdaeme}, {Nuza}, {Olmstead}, {Oravetz}, {Oravetz},
  {Owen}, {Padmanabhan}, {Palanque-Delabrouille}, {Pan}, {Parejko},
  {P{\^a}ris}, {Percival}, {P{\'e}rez-Fournon}, {P{\'e}rez-R{\`a}fols},
  {Petitjean}, {Pfaffenberger}, {Pforr}, {Pieri}, {Prada}, {Price-Whelan},
  {Raddick}, {Rebolo}, {Rich}, {Richards}, {Rockosi}, {Roe}, {Ross}, {Ross},
  {Rossi}, {Rubi{\~n}o-Martin}, {Samushia}, {S{\'a}nchez}, {Sayres}, {Schmidt},
  {Schneider}, {Sc{\'o}ccola}, {Seo}, {Shelden}, {Sheldon}, {Shen}, {Shu},
  {Slosar}, {Smee}, {Snedden}, {Stauffer}, {Steele}, {Strauss}, {Streblyanska},
  {Suzuki}, {Swanson}, {Tal}, {Tanaka}, {Thomas}, {Tinker}, {Tojeiro},
  {Tremonti}, {Vargas Maga{\~n}a}, {Verde}, {Viel}, {Wake}, {Watson}, {Weaver},
  {Weinberg}, {Weiner}, {West}, {White}, {Wood-Vasey}, {Yeche}, {Zehavi},
  {Zhao}, \& {Zheng}}]{Dawson13}
{Dawson}, K.~S., {et~al.} 2013, \aj, 145, 10

\bibitem[{{Dere} {et~al.}(1997){Dere}, {Landi}, {Mason}, {Monsignori Fossi}, \&
  {Young}}]{Dere97}
{Dere}, K.~P., {Landi}, E., {Mason}, H.~E., {Monsignori Fossi}, B.~C., \&
  {Young}, P.~R. 1997, \aaps, 125, 149

\bibitem[{{Dexter} \& {Agol}(2011)}]{Dexter11}
{Dexter}, J., \& {Agol}, E. 2011, \apjl, 727, L24

\bibitem[{{Di~Matteo} {et~al.}(2005){Di~Matteo}, {Springel}, \&
  {Hernquist}}]{Dimatteo05}
{Di~Matteo}, T., {Springel}, V., \& {Hernquist}, L. 2005, \nat, 433, 604

\bibitem[{{Dunn} {et~al.}(2010){Dunn}, {Bautista}, {Arav}, {Moe}, {Korista},
  {Costantini}, {Benn}, {Ellison}, \& {Edmonds}}]{Dunn10}
{Dunn}, J.~P., {et~al.} 2010, \apj, 709, 611

\bibitem[{{Eisenstein} {et~al.}(2011){Eisenstein}, {Weinberg}, {Agol},
  {Aihara}, {Allende Prieto}, {Anderson}, {Arns}, {Aubourg}, {Bailey},
  {Balbinot}, \& et~al.}]{Eisenstein11}
{Eisenstein}, D.~J., {et~al.} 2011, \aj, 142, 72

\bibitem[{{Faucher-Gigu{\`e}re} {et~al.}(2012){Faucher-Gigu{\`e}re},
  {Quataert}, \& {Murray}}]{Faucher12}
{Faucher-Gigu{\`e}re}, C.-A., {Quataert}, E., \& {Murray}, N. 2012, \mnras,
  420, 1347

\bibitem[{{Filiz Ak} {et~al.}(2012){Filiz Ak}, {Brandt}, {Hall}, {Schneider},
  {Anderson}, {Gibson}, {Lundgren}, {Myers}, {Petitjean}, {Ross}, {Shen},
  {York}, {Bizyaev}, {Brinkmann}, {Malanushenko}, {Oravetz}, {Pan}, {Simmons},
  \& {Weaver}}]{Filizak12}
{Filiz Ak}, N., {et~al.} 2012, \apj, 757, 114

\bibitem[{{Filiz Ak} {et~al.}(2013){Filiz Ak}, {Brandt}, {Hall}, {Schneider},
  {Anderson}, {Hamann}, {Lundgren}, {Myers}, {P{\^a}ris}, {Petitjean}, {Ross},
  {Shen}, \& {York}}]{Filizak13}
---. 2013, \apj, 777, 168

\bibitem[{{Fukugita} {et~al.}(1996){Fukugita}, {Ichikawa}, {Gunn}, {Doi},
  {Shimasaku}, \& {Schneider}}]{Fukugita96}
{Fukugita}, M., {Ichikawa}, T., {Gunn}, J.~E., {Doi}, M., {Shimasaku}, K., \&
  {Schneider}, D.~P. 1996, \aj, 111, 1748

\bibitem[{{Gallagher} {et~al.}(2004){Gallagher}, {Brandt}, {Wills}, {Charlton},
  {Chartas}, \& {Laor}}]{Gallagher04}
{Gallagher}, S.~C., {Brandt}, W.~N., {Wills}, B.~J., {Charlton}, J.~C.,
  {Chartas}, G., \& {Laor}, A. 2004, \apj, 603, 425

\bibitem[{{Gibson} \& {Brandt}(2012)}]{Gibson12}
{Gibson}, R.~R., \& {Brandt}, W.~N. 2012, \apj, 746, 54

\bibitem[{{Gibson} {et~al.}(2010){Gibson}, {Brandt}, {Gallagher}, {Hewett}, \&
  {Schneider}}]{Gibson10}
{Gibson}, R.~R., {Brandt}, W.~N., {Gallagher}, S.~C., {Hewett}, P.~C., \&
  {Schneider}, D.~P. 2010, \apj, 713, 220

\bibitem[{{Gibson} {et~al.}(2008){Gibson}, {Brandt}, \& {Schneider}}]{Gibson08}
{Gibson}, R.~R., {Brandt}, W.~N., \& {Schneider}, D.~P. 2008, \apj, 685, 773

\bibitem[{{Gibson} {et~al.}(2009){Gibson}, {Jiang}, {Brandt}, {Hall}, {Shen},
  {Wu}, {Anderson}, {Schneider}, {Vanden Berk}, {Gallagher}, {Fan}, \&
  {York}}]{Gibson09b}
{Gibson}, R.~R., {et~al.} 2009, \apj, 692, 758

\bibitem[{{Gunn} {et~al.}(2006){Gunn}, {Siegmund}, {Mannery}, {Owen}, {Hull},
  {Leger}, {Carey}, {Knapp}, {York}, {Boroski}, {Kent}, {Lupton}, {Rockosi},
  {Evans}, {Waddell}, {Anderson}, {Annis}, {Barentine}, {Bartoszek}, {Bastian},
  {Bracker}, {Brewington}, {Briegel}, {Brinkmann}, {Brown}, {Carr},
  {Czarapata}, {Drennan}, {Dombeck}, {Federwitz}, {Gillespie}, {Gonzales},
  {Hansen}, {Harvanek}, {Hayes}, {Jordan}, {Kinney}, {Klaene}, {Kleinman},
  {Kron}, {Kresinski}, {Lee}, {Limmongkol}, {Lindenmeyer}, {Long}, {Loomis},
  {McGehee}, {Mantsch}, {Neilsen}, {Neswold}, {Newman}, {Nitta}, {Peoples},
  {Pier}, {Prieto}, {Prosapio}, {Rivetta}, {Schneider}, {Snedden}, \&
  {Wang}}]{Gunn06}
{Gunn}, J.~E., {et~al.} 2006, \aj, 131, 2332

\bibitem[{{Gwyn}(2008)}]{Gwyn08}
{Gwyn}, S.~D.~J. 2008, \pasp, 120, 212

\bibitem[{{Hall} {et~al.}(2011){Hall}, {Anosov}, {White}, {Brandt}, {Gregg},
  {Gibson}, {Becker}, \& {Schneider}}]{Hall11}
{Hall}, P.~B., {Anosov}, K., {White}, R.~L., {Brandt}, W.~N., {Gregg}, M.~D.,
  {Gibson}, R.~R., {Becker}, R.~H., \& {Schneider}, D.~P. 2011, \mnras, 411,
  2653
  
\bibitem[{{Hall} {et~al.}(2003){Hall}, {Hutsemekers}, {Anderson}, {Brinkmann}, {Fan},
  {Schneider},  \& {York}}]{Hall03}
{Hall}, P.~B., {Hutsemekers}, D., {Anderson}, S.~F., {Brinkmann}, J., {Fan}, X.,
  {Schneider}, D.~P., \& {York}, D.~G 2003, \apj, 593, 189


\bibitem[{{Hamann} {et~al.}(1995){Hamann}, {Barlow}, {Beaver}, {Burbidge},
  {Cohen}, {Junkkarinen}, \& {Lyons}}]{Hamann95}
{Hamann}, F., {Barlow}, T.~A., {Beaver}, E.~A., {Burbidge}, E.~M., {Cohen},
  R.~D., {Junkkarinen}, V., \& {Lyons}, R. 1995, \apj, 443, 606

\bibitem[{{Higginbottom} {et~al.}(2013){Higginbottom}, {Knigge}, {Long}, {Sim},
  \& {Matthews}}]{Higginbottom13}
{Higginbottom}, N., {Knigge}, C., {Long}, K.~S., {Sim}, S.~A., \& {Matthews},
  J.~H. 2013, \mnras, 436, 1390

\bibitem[{{Hopkins} \& {Elvis}(2010)}]{Hopkins10}
{Hopkins}, P.~F., \& {Elvis}, M. 2010, \mnras, 401, 7

\bibitem[{{Hopkins} {et~al.}(2005){Hopkins}, {Hernquist}, {Cox}, {Di Matteo},
  {Martini}, {Robertson}, \& {Springel}}]{Hopkins05a}
{Hopkins}, P.~F., {Hernquist}, L., {Cox}, T.~J., {Di Matteo}, T., {Martini},
  P., {Robertson}, B., \& {Springel}, V. 2005, \apj, 630, 705

\bibitem[{{Kallman} \& {McCray}(1982)}]{Kallman82}
{Kallman}, T.~R., \& {McCray}, R. 1982, \apjs, 50, 263

\bibitem[{{Kaspi} {et~al.}(2007){Kaspi}, {Brandt}, {Maoz}, {Netzer},
  {Schneider}, \& {Shemmer}}]{Kaspi07}
{Kaspi}, S., {Brandt}, W.~N., {Maoz}, D., {Netzer}, H., {Schneider}, D.~P., \&
  {Shemmer}, O. 2007, \apj, 659, 997

\bibitem[{{Kelly}(2007)}]{Kelly07b}
{Kelly}, B.~C. 2007, \apj, 665, 1489

\bibitem[{{King}(2010)}]{King10}
{King}, A.~R. 2010, \mnras, 402, 1516

\bibitem[{{Konigl} \& {Kartje}(1994)}]{Konigl94}
{Konigl}, A., \& {Kartje}, J.~F. 1994, \apj, 434, 446

\bibitem[{{Krolik}(1999)}]{Krolik99}
{Krolik}, J.~H. 1999, {Active galactic nuclei : from the central black hole to
  the galactic environment}

\bibitem[{{Krolik} \& {Kriss}(1995)}]{Krolik95}
{Krolik}, J.~H., \& {Kriss}, G.~A. 1995, \apj, 447, 512

\bibitem[{{Landi} {et~al.}(2013){Landi}, {Young}, {Dere}, {Del Zanna}, \&
  {Mason}}]{Landi13}
{Landi}, E., {Young}, P.~R., {Dere}, K.~P., {Del Zanna}, G., \& {Mason}, H.~E.
  2013, \apj, 763, 86

\bibitem[{{Laor} \& {Brandt}(2002)}]{Laor02}
{Laor}, A., \& {Brandt}, W.~N. 2002, \apj, 569, 641

\bibitem[{{Leighly} {et~al.}(2009){Leighly}, {Hamann}, {Casebeer}, \&
  {Grupe}}]{Leighly09}
{Leighly}, K.~M., {Hamann}, F., {Casebeer}, D.~A., \& {Grupe}, D. 2009, \apj,
  701, 176

\bibitem[{{Lucy} {et~al.}(2014){Lucy}, {Leighly}, {Terndrup}, {Dietrich}, \&
  {Gallagher}}]{Lucy14}
{Lucy}, A.~B., {Leighly}, K.~M., {Terndrup}, D.~M., {Dietrich}, M., \&
  {Gallagher}, S.~C. 2014, \apj, 783, 58

\bibitem[{{Lundgren} {et~al.}(2007){Lundgren}, {Wilhite}, {Brunner}, {Hall},
  {Schneider}, {York}, {Vanden Berk}, \& {Brinkmann}}]{Lundgren07}
{Lundgren}, B.~F., {Wilhite}, B.~C., {Brunner}, R.~J., {Hall}, P.~B.,
  {Schneider}, D.~P., {York}, D.~G., {Vanden Berk}, D.~E., \& {Brinkmann}, J.
  2007, \apj, 656, 73

\bibitem[{{Marshall} {et~al.}(1996){Marshall}, {Carone}, {Shull}, {Malkan}, \&
  {Elvis}}]{Marshall96}
{Marshall}, H.~L., {Carone}, T.~E., {Shull}, J.~M., {Malkan}, M.~A., \&
  {Elvis}, M. 1996, \apj, 457, 169

\bibitem[{{Marshall} {et~al.}(1997){Marshall}, {Carone}, {Peterson}, {Clavel},
  {Crenshaw}, {Korista}, {Kriss}, {Krolik}, {Malkan}, {Morris}, \&
  {Reichert}}]{Marshall97}
{Marshall}, H.~L., {et~al.} 1997, \apj, 479, 222

\bibitem[{{Moll} {et~al.}(2007){Moll}, {Schindler}, {Domainko}, {Kapferer},
  {Mair}, {van Kampen}, {Kronberger}, {Kimeswenger}, \& {Ruffert}}]{Moll07}
{Moll}, R., {et~al.} 2007, \aap, 463, 513

\bibitem[{{Murray} {et~al.}(1995){Murray}, {Chiang}, {Grossman}, \&
  {Voit}}]{Murray95}
{Murray}, N., {Chiang}, J., {Grossman}, S.~A., \& {Voit}, G.~M. 1995, \apj,
  451, 498

\bibitem[{{P{\^a}ris} {et~al.}(2012){P{\^a}ris}, {Petitjean}, {Aubourg},
  {Bailey}, {Ross}, {Myers}, {Strauss}, {Anderson}, {Arnau}, {Bautista},
  {Bizyaev}, {Bolton}, {Bovy}, {Brandt}, {Brewington}, {Browstein}, {Busca},
  {Capellupo}, {Carithers}, {Croft}, {Dawson}, {Delubac}, {Ebelke},
  {Eisenstein}, {Engelke}, {Fan}, {Filiz Ak}, {Finley}, {Font-Ribera}, {Ge},
  {Gibson}, {Hall}, {Hamann}, {Hennawi}, {Ho}, {Hogg}, {Ivezi{\'c}}, {Jiang},
  {Kimball}, {Kirkby}, {Kirkpatrick}, {Lee}, {Le Goff}, {Lundgren}, {MacLeod},
  {Malanushenko}, {Malanushenko}, {Maraston}, {McGreer}, {McMahon},
  {Miralda-Escud{\'e}}, {Muna}, {Noterdaeme}, {Oravetz},
  {Palanque-Delabrouille}, {Pan}, {Perez-Fournon}, {Pieri}, {Richards},
  {Rollinde}, {Sheldon}, {Schlegel}, {Schneider}, {Slosar}, {Shelden}, {Shen},
  {Simmons}, {Snedden}, {Suzuki}, {Tinker}, {Viel}, {Weaver}, {Weinberg},
  {White}, {Wood-Vasey}, \& {Y{\`e}che}}]{Paris12}
{P{\^a}ris}, I., {et~al.} 2012, \aap, 548, A66

\bibitem[{{P{\^a}ris} {et~al.}(2014){P{\^a}ris}, {Petitjean}, {Aubourg},
  {Ross}, {Myers}, {Streblyanska}, {Bailey}, {Hall}, {Strauss}, {Anderson},
  {Bizyaev}, {Borde}, {Brinkmann}, {Bovy}, {Brandt}, {Brewington},
  {Brownstein}, {Cook}, {Ebelke}, {Fan}, {Filiz Ak}, {Finley}, {Font-Ribera},
  {Ge}, {Hamann}, {Ho}, {Jiang}, {Kinemuchi}, {Malanushenko}, {Malanushenko},
  {Marchante}, {McGreer}, {McMahon}, {Miralda-Escud{\'e}}, {Muna},
  {Noterdaeme}, {Oravetz}, {Palanque-Delabrouille}, {Pan}, {Perez-Fournon},
  {Pieri}, {Riffel}, {Schlegel}, {Schneider}, {Simmons}, {Viel}, {Weaver},
  {Wood-Vasey}, {Y{\`e}che}, \& {York}}]{Paris14}
---. 2014, \aap, 563, A54

\bibitem[{{P{\^a}ris} {et~al.}(2015){P{\^a}ris}, {Petitjean}, {Aubourg},
  {Ross}, {Myers}, {Streblyanska}, {Bailey}, {Hall}, {Strauss}, {Anderson},
  {Bizyaev}, {Borde}, {Brinkmann}, {Bovy}, {Brandt}, {Brewington},
  {Brownstein}, {Cook}, {Ebelke}, {Fan}, {Filiz Ak}, {Finley}, {Font-Ribera},
  {Ge}, {Hamann}, {Ho}, {Jiang}, {Kinemuchi}, {Malanushenko}, {Malanushenko},
  {Marchante}, {McGreer}, {McMahon}, {Miralda-Escud{\'e}}, {Muna},
  {Noterdaeme}, {Oravetz}, {Palanque-Delabrouille}, {Pan}, {Perez-Fournon},
  {Pieri}, {Riffel}, {Schlegel}, {Schneider}, {Simmons}, {Viel}, {Weaver},
  {Wood-Vasey}, {Y{\`e}che}, \& {York}}]{Paris15}
---. 2015, \aap, in preparation

\bibitem[{{Pei}(1992)}]{Pei92}
{Pei}, Y.~C. 1992, \apj, 395, 130

\bibitem[{{Peterson} {et~al.}(1998){Peterson}, {Wanders}, {Horne}, {Collier},
  {Alexander}, {Kaspi}, \& {Maoz}}]{Peterson98b}
{Peterson}, B.~M., {Wanders}, I., {Horne}, K., {Collier}, S., {Alexander}, T.,
  {Kaspi}, S., \& {Maoz}, D. 1998, \pasp, 110, 660

\bibitem[{{Peterson} {et~al.}(2000){Peterson}, {McHardy}, {Wilkes}, {Berlind},
  {Bertram}, {Calkins}, {Collier}, {Huchra}, {Mathur}, {Papadakis}, {Peters},
  {Pogge}, {Romano}, {Tokarz}, {Uttley}, {Vestergaard}, \&
  {Wagner}}]{Peterson00b}
{Peterson}, B.~M., {et~al.} 2000, \apj, 542, 161

\bibitem[{{Proga} {et~al.}(2000){Proga}, {Stone}, \& {Kallman}}]{Proga00}
{Proga}, D., {Stone}, J.~M., \& {Kallman}, T.~R. 2000, \apj, 543, 686

\bibitem[{{Rafiee} \& {Hall}(2011)}]{Rafiee11}
{Rafiee}, A., \& {Hall}, P.~B. 2011, \apjs, 194, 42

\bibitem[{{Richards} {et~al.}(2006){Richards}, {Lacy}, {Storrie-Lombardi},
  {Hall}, {Gallagher}, {Hines}, {Fan}, {Papovich}, {Vanden Berk}, {Trammell},
  {Schneider}, {Vestergaard}, {York}, {Jester}, {Anderson}, {Budav{\'a}ri}, \&
  {Szalay}}]{Richards06}
{Richards}, G.~T., {et~al.} 2006, \apjs, 166, 470

\bibitem[{{Richards} {et~al.}(2011){Richards}, {Kruczek}, {Gallagher}, {Hall},
  {Hewett}, {Leighly}, {Deo}, {Kratzer}, \& {Shen}}]{Richards11}
---. 2011, \aj, 141, 167

\bibitem[{{Rogerson} {et~al.}(2011){Rogerson}, {Hall}, {Snedden}, {Brotherton},
  \& {Anderson}}]{Rogerson11}
{Rogerson}, J.~A., {Hall}, P.~B., {Snedden}, S.~A., {Brotherton}, M.~S., \&
  {Anderson}, S.~F. 2011, New Astron., 16, 128

\bibitem[{{Saez} {et~al.}(2012){Saez}, {Brandt}, {Gallagher}, {Bauer}, \&
  {Garmire}}]{Saez12}
{Saez}, C., {Brandt}, W.~N., {Gallagher}, S.~C., {Bauer}, F.~E., \& {Garmire},
  G.~P. 2012, \apj, 759, 42

\bibitem[{{Schlegel} {et~al.}(1998){Schlegel}, {Finkbeiner}, \&
  {Davis}}]{Schlegel98}
{Schlegel}, D.~J., {Finkbeiner}, D.~P., \& {Davis}, M. 1998, \apj, 500, 525

\bibitem[{{Shemmer} {et~al.}(2014){Shemmer}, {Brandt}, {Paolillo}, {Kaspi},
  {Vignali}, {Stein}, {Lira}, {Schneider}, \& {Gibson}}]{Shemmer14}
{Shemmer}, O., {et~al.} 2014, \apj, 783, 116

\bibitem[{{Shen} {et~al.}(2015){Shen}, {Brandt}, {Dawson}, {Hall}, {McGreer},
  {Anderson}, {Chen}, {Denney}, {Eftekharzadeh}, {Fan}, {Gao}, {Green},
  {Greene}, {Ho}, {Horne}, {Jiang}, {Kelly}, {Kinemuchi}, {Kochanek},
  {P{\^a}ris}, {Peters}, {Peterson}, {Petitjean}, {Ponder}, {Richards},
  {Schneider}, {Seth}, {Smith}, {Strauss}, {Tao}, {Trump}, {Wood-Vasey}, {Zu},
  {Eisenstein}, {Pan}, {Bizyaev}, {Malanushenko}, {Malanushenko}, \&
  {Oravetz}}]{Shen15}
{Shen}, Y., {et~al.} 2015, \apjs, 216, 4

\bibitem[{{Sim} {et~al.}(2010){Sim}, {Proga}, {Miller}, {Long}, \&
  {Turner}}]{Sim10}
{Sim}, S.~A., {Proga}, D., {Miller}, L., {Long}, K.~S., \& {Turner}, T.~J.
  2010, \mnras, 408, 1396

\bibitem[{{Smee} {et~al.}(2013){Smee}, {Gunn}, {Uomoto}, {Roe}, {Schlegel},
  {Rockosi}, {Carr}, {Leger}, {Dawson}, {Olmstead}, {Brinkmann}, {Owen},
  {Barkhouser}, {Honscheid}, {Harding}, {Long}, {Lupton}, {Loomis}, {Anderson},
  {Annis}, {Bernardi}, {Bhardwaj}, {Bizyaev}, {Bolton}, {Brewington}, {Briggs},
  {Burles}, {Burns}, {Castander}, {Connolly}, {Davenport}, {Ebelke}, {Epps},
  {Feldman}, {Friedman}, {Frieman}, {Heckman}, {Hull}, {Knapp}, {Lawrence},
  {Loveday}, {Mannery}, {Malanushenko}, {Malanushenko}, {Merrelli}, {Muna},
  {Newman}, {Nichol}, {Oravetz}, {Pan}, {Pope}, {Ricketts}, {Shelden},
  {Sandford}, {Siegmund}, {Simmons}, {Smith}, {Snedden}, {Schneider},
  {SubbaRao}, {Tremonti}, {Waddell}, \& {York}}]{Smee13}
{Smee}, S.~A., {et~al.} 2013, \aj, 146, 32

\bibitem[{{Springel} {et~al.}(2005){Springel}, {Di~Matteo}, \&
  {Hernquist}}]{Springel05b}
{Springel}, V., {Di~Matteo}, T., \& {Hernquist}, L. 2005, \mnras, 361, 776

\bibitem[{{Stepanian} {et~al.}(1998){Stepanian}, {Lipovetsky}, {Chavushyan},
  {Erastova}, \& {Balayan}}]{Stepanian98}
{Stepanian}, J.~A., {Lipovetsky}, V.~A., {Chavushyan}, V.~H., {Erastova},
  L.~K., \& {Balayan}, S.~K. 1998, ArXiv Astrophysics e-prints

\bibitem[{{Vanden Berk} {et~al.}(2001){Vanden Berk}, {Richards}, {Bauer},
  {Strauss}, {Schneider}, {Heckman}, {York}, {Hall}, {Fan}, {Knapp},
  {Anderson}, {Annis}, {Bahcall}, {Bernardi}, {Briggs}, {Brinkmann}, {Brunner},
  {Burles}, {Carey}, {Castander}, {Connolly}, {Crocker}, {Csabai}, {Doi},
  {Finkbeiner}, {Friedman}, {Frieman}, {Fukugita}, {Gunn}, {Hennessy},
  {Ivezi{\'c}}, {Kent}, {Kunszt}, {Lamb}, {Leger}, {Long}, {Loveday}, {Lupton},
  {Meiksin}, {Merelli}, {Munn}, {Newberg}, {Newcomb}, {Nichol}, {Owen}, {Pier},
  {Pope}, {Rockosi}, {Schlegel}, {Siegmund}, {Smee}, {Snir}, {Stoughton},
  {Stubbs}, {SubbaRao}, {Szalay}, {Szokoly}, {Tremonti}, {Uomoto}, {Waddell},
  {Yanny}, \& {Zheng}}]{Vandenberk01}
{Vanden Berk}, D.~E., {et~al.} 2001, \aj, 122, 549

\bibitem[{{Vivek} {et~al.}(2012){Vivek}, {Srianand}, {Mahabal}, \&
  {Kuriakose}}]{Vivek12}
{Vivek}, M., {Srianand}, R., {Mahabal}, A., \& {Kuriakose}, V.~C. 2012, \mnras,
  421, L107

\bibitem[{{Weymann} {et~al.}(1991){Weymann}, {Morris}, {Foltz}, \&
  {Hewett}}]{Weymann91}
{Weymann}, R.~J., {Morris}, S.~L., {Foltz}, C.~B., \& {Hewett}, P.~C. 1991,
  \apj, 373, 23

\end{thebibliography}

\end{document}